\documentclass[sigconf]{acmart}
\usepackage{booktabs}
\usepackage{tikz}
\usepackage{wrapfig}
\usepackage{array}
\usepackage{subcaption}
\usepackage{amsmath}
\usepackage{multirow}
\usepackage{amsthm}
\usepackage{dsfont}
\usepackage{newfloat}
\usepackage{listings}
\usepackage[norelsize, linesnumbered, ruled, lined, boxed, commentsnumbered]{algorithm2e}
\usepackage[switch]{lineno}

\newcommand{\ignore}[1]{}

\AtBeginDocument{%
  }

\setcopyright{acmlicensed}
\copyrightyear{2026}
\acmYear{2026}
\acmDOI{XXXXXXX.XXXXXXX}
\acmConference[Under Review, KDD '26]{}{Aug 9-13,26}{Jeju, Korea}
\acmISBN{978-1-4503-XXXX-X/2018/06}

\begin{document}

\title{Emergence of Phase Transitions in Complex Contagions}
\author{Saurabh Sharma}
\affiliation{%
  \institution{University of California, Santa Barbara}
  \country{United States}
}
 \email{saurabhsharma@ucsb.edu}

\author{Ambuj Singh}
\affiliation{%
  \institution{University of California, Santa Barbara}
  \country{United States}
}
 \email{ambuj@cs.ucsb.edu}

\renewcommand{\shortauthors}{Saurabh Sharma \& Ambuj Singh}

\begin{abstract}
Understanding how complex behaviors, opinions, and innovations spread in online social networks remains a central challenge in computational social science. Existing models of complex contagion typically rely on stylized threshold mechanisms based solely on the number of infected neighbors and do not account for the interaction between individual preferences, local social influence, and global sentiment. Moreover, the emergence of virality through phase transitions and tipping points remains poorly characterized.

In this paper, we propose a unified propagation cascade model in which notions propagate as high-dimensional vectors in the same feature space as network nodes. Node activations are governed by a unified decision function that integrates propagation affinity, local influence, and global influence. The resulting dynamics induce a stochastic, Markovian cascade process that enables efficient MCMC sampling of propagation outcomes.

Using preferential attachment networks, we systematically study spread distributions, incubation dynamics, parameter sensitivity, and phase transition behavior. Our results show that balanced interactions between local reinforcement and global activation are critical for successful cascades and that early-stage growth patterns provide reliable signals of impending phase transitions.
\end{abstract}



\maketitle

\section{Introduction}
Complex contagion \cite{centola2007complex,ebrahimi2017complex,guilbeault2018complex,centola2021influencers} is a popular model for the spread of behaviours in a networked system where nodes adopt the propagating behaviour if and only if the cumulative local influence exceeds a threshold. In contrast to phenomena such as information and disease spread, the spreading of behaviours is slower and requires critical mass to gather momentum. However, most prior models of complex contagion rely solely on the number of infected neighbors as a measure of local influence. This is oversimplified and does not easily fit into the world of online social networks (OSN) where nodes keep public profiles and adoption depends not only on a single summary statistic, but on the complex interaction between three factors (1) the propagating notion itself, (2) the properties of the local influencing neighbors, and (3) the global sentiment towards adoption (bullish/bearish) that is made available by the OSN provider. Further, viral memes usually demonstrate a phase transition phenomenon where the meme initially grows slowly and then rapidly accelerates towards virality. This phase transition is insufficiently understood by prior work that concentrates on the time to convergence but not on the "tipping point" itself. In this work, we go a step further and propose a model that accounts for both the complex OSN interplays and the phase transition phenomenon. 

To model OSN dynamics, we postulate a Unified Propagation (UP) model that has the following salient properties, (1) Propagations are modelled as vectors in the same high-dimensional space where node features exist,  (2) Propagations spread in a synchronous stepwise manner whereby network states are updated simulataneously at discrete time intervals, (3) Nodes activate based on a shared decision function with shared thresholds that accounts for propagation affinity, local influence and global influence, and (4) The underlying network is based on the Preferential Attachment (PA) model ~\cite{barabasi1999emergence} which yields a core-periphery structure and a scale-free degree distribution observed in real-world networks. Our model affords sampling ease from the propagation distribution which is similar to a gated, time-varying random walk on a network.

The design of our propagation model is motivated by recent work at the intersection of computation and social science \cite{centola2021influencers} that shows small topic-oriented peripheral groups can have a large influence through network propagation pathways from periphery to core. Moreover, such groups are essential for incubating new concepts in a controlled environment, from where they can spread to large portions of the population through core-periphery pathways. Further, they report cases studies such as Google Glass, where propagations from a core node die out quickly when initial adoption is slow, which signals a population wide unpopularity that impedes adoption. Therefore, both the initial incubation and global influence factors are necessary for determining the spread of the complex contagion. Finally, as we show in our experiments, we can observe the emergence of phase transitions in complex networked systems \cite{shrager1987observation,scheffer2012anticipating} by running simulations on our proposed model.

Our framework leverages propagation sampling from the high-dimensional node feature space to yield a simple personal affinity score. Prior works in psycho-social belief dynamics \cite{galesic2021integrating} usually incorporate latent personal factors with publicly visible attributes to determine individual and consequently collective behaviour. Our approach is closest to \cite{gaitonde2021polarization} who model node opinions as unit-length vectors and propagate randomly sampled opinion vectors based on correlation scores. However, we differenciate ourselves by applying this approach to network cascades, which are distinct from opinion dynamics models in that they're stochastically spreading activations on a network. To the best of our knowledge, we're the first to model propagations as vectors for network cascades. 




Our main contributions are summarized as follows:
\begin{itemize}
    \item We introduce a general propagation-based model of complex contagion in which propagating notions and node attributes are represented in a shared high-dimensional feature space.
    \item We propose the Unified Propagation (UP) model that integrates propagation affinity, local influence, and global influence, yielding a stochastic Markovian cascade process.
    \item We develop an MCMC-based simulation framework for sampling from the distribution of cascade realizations induced by the model.
    \item We analyze incubation dynamics and derive conditions under which local reinforcement can lead to global diffusion through core--periphery pathways.
    \item We empirically demonstrate phase transition behavior and tipping points in cascade growth and show how early-stage dynamics can be used to anticipate virality.
\end{itemize}

\section{Related Work}

\paragraph{\textbf{Diffusion of innovations.}}
The study of social diffusion has a long tradition in sociology, where adoption is understood as a socially embedded process shaped by interpersonal influence, heterogeneity in individual thresholds, and the structure of communication networks. Early work on diffusion of innovations emphasizes the role of opinion leaders, social systems, and communication channels in shaping adoption trajectories \cite{rogers2003diffusion}. Complementary threshold-based models describe how collective behavior can emerge from heterogeneous individual decision rules driven by local social exposure \cite{granovetter1978threshold}. Empirical and theoretical studies in this line highlight that adoption is rarely driven by single exposures alone, but instead depends on cumulative social reinforcement, normative pressures, and perceived legitimacy within social groups. This sociological perspective motivates the need for models that explicitly account for local context and heterogeneous influence pathways, rather than relying solely on aggregate exposure counts.
\paragraph{\textbf{Complex contagion and computational models.}}
Building on these foundations, the notion of complex contagion formalizes diffusion processes in which multiple reinforcing contacts are required before adoption occurs \cite{centola2007complex}. Computational and algorithmic models have investigated how network topology, clustering, and community structure affect the emergence of large-scale cascades under such reinforcement mechanisms \cite{watts2002simple,kleinberg2007cascading}. Subsequent work has studied algorithmic and structural properties of complex contagions, including the role of strong ties and wide bridges, the effect of network interventions, and the computational aspects of triggering global cascades under threshold dynamics \cite{ghasemiesfeh2013complex,ebrahimi2015complex}. Our work further emphasizes that reinforcement-based spreading processes exhibit sharp changes in behavior depending on network structure and initial conditions, suggesting the presence of critical regimes that separate small, localized cascades from system-wide diffusion.
\paragraph{\textbf{Network propagation, label propagation, and diffusion processes.}}
A complementary line of work models propagation through linear diffusion mechanisms such as random walks, label propagation, and heat or diffusion kernels. Seminal work by Zhu and Ghahramani formulates semi-supervised learning as a harmonic extension problem on graphs, in which labels propagate via random walks with absorbing boundary conditions \cite{zhu2002learning,zhu2003semi}. Related formulations based on graph Laplacians and heat kernels characterize propagation through continuous-time diffusion processes and spectral filtering on networks \cite{zhou2004learning,kondor2002diffusion}. These methods yield closed-form solutions and strong convergence guarantees under mild connectivity assumptions. While such models have primarily been used for inference and representation learning, their random-walk interpretation provides a natural foundation for sampling-based propagation mechanisms. Our work builds on this perspective by coupling propagation in a high-dimensional feature space with stochastic activation dynamics, thereby extending linear diffusion models toward the sampling of network cascades.
\paragraph{\textbf{Network formation and structural models.}}
The behavior of diffusion and cascade processes is tightly linked to the generative mechanisms underlying network structure. Classical models of network formation include the small-world model, which captures short path lengths and high clustering through random rewiring \cite{watts1998collective}, and preferential attachment models that produce scale-free degree distributions and pronounced core--periphery structure \cite{barabasi1999emergence}. In addition, Kleinberg's navigable small-world model introduces a geometric formulation that explains how long-range links enable efficient decentralized routing and shape information flow patterns \cite{kleinberg2000navigation}. These models have been widely used as benchmarks for studying diffusion, contagion, and search processes, and they provide principled abstractions for the heterogeneous connectivity and structural bottlenecks observed in online social networks. In this work, we adopt a preferential attachment framework to explicitly capture the emergence of core--periphery pathways that play a central role in our propagation dynamics.
\paragraph{\textbf{Phase transitions and virality in networked systems.}}
Diffusion and contagion processes on networks are known to exhibit phase transitions in which small changes in parameters or initial conditions lead to abrupt changes in macroscopic behavior. Early studies in complex systems and statistical physics have characterized critical points and tipping behavior in interacting systems \cite{scheffer2012anticipating}. Likewise, models of spreading activation networks in psychological theories of cognition~\cite{shrager1987observation} have demonstrated phase transitions by changing network topology or the activation parameters across phase boundaries. In the context of social and information networks, recent work has investigated how viral cascades transition from slow growth to rapid, large-scale spread, and how structural properties of cascade trees relate to different notions of virality \cite{goel2015structural}. Related studies on epidemic and information spreading further show that network heterogeneity and degree distributions fundamentally affect critical thresholds and outbreak dynamics \cite{pastorsatorras2001epidemic}. These results motivate our focus on the transient evolution of cascades and the emergence of tipping points, rather than solely on asymptotic convergence behavior.

\section{Method}
\subsection{Preliminaries}
Consider a graph $G=(A,X)$, with a weighted adjacency matrix $A \in [0,1]^{n\times n}$ and node attribute matrix $X \in [0,1]^{n\times d}$ respectively, and node activations $Y \in \{0,1\}^n$. We refer to the associated node-ids as $\mathcal{V}=\{1,\dots,n\}$. We denote the node feature $x_v \in [0,1]^{d}$, and the node activation $y_v \in \{0,1\}$. The active set of nodes is denoted as $S = \{v|y_v=1\}$. 
Further, consider a propagation $C \in [0,1]^d$ initiated at seed set $S_0$, creating a cascade of network activations $Y^t$ over time steps $t=1,\dots,T_{P,S_0}$,  with $T_{C,S_0}$ denoting the convergence time. The spreading dynamics defined by the propagation and activation rules form a discrete-time Markov process, since the network state at time $t$ depends only on the state at time $t-1$ and $C$.

The networks we use in the paper are based on the Preferential Attachment (PA) model ~\cite{barabasi1999emergence} of network growth and link formation. The PA model constructs networks incrementally, unlike the configuration model ~\cite{newman2018networks} or the Watts-Strogatz small world model ~\cite{watts1998collective}, and matches the power-law degree distribution of real world networks by construction. Formally, a PA graph $G(n,r)$ is a network of $n$ nodes where each incoming node attaches to $r$ existing nodes according to a probability proportional to their degree. Therefore, this creates an early mover advantage, causing initial nodes to have high degree centerality and later nodes to be on the periphery. The resulting core-periphery network structure is ideal for our analysis of the spread of complex contagions. Note that PA networks have been used by prior work ~\cite{ebrahimi2017complex}, but our contagion model exploits the core-periphery structure more explicitly and gives power to the periphery as expected from sociological analysis ~\cite{centola2021influencers}.   

Node features are constructed using Laplacian node embeddings, in other words, by taking the $k$ smallest eigenvectors of the graph Laplacian. Formally, the graph Laplacian $L=D-A$, where $D$ is the degree matrix of the graph, with its eigenvector decomposition given by $L=Q\Delta Q^T$ as usual. Therefore, the node feature matrix is the submatrix $X = Q_{:,:k}$. The graph Laplacian appears frequently in graph signal processing, with the eigenvalues-eigenvector pairs yielding frequencies in the graph Fourier domain. The smallest eigenvectors correspond to low frequencies that vary smoothly over the graph and carry structural information as well. Moreover, certain eigenvectors carry important information about the connectivity of the graph ~\cite{spielman2012spectral} and are used in graph cuts, community detection and clustering ~\cite{shi2000normalized}. Therefore, while there exist other node embedding approaches ~\cite{perozzi2014deepwalk,grover2016node2vec} that have shown promise in problems such as node classification and link detection, we use Laplacian embeddings in this work due to their efficient construction using eigenvector factorization.    

Since the PA model yields an unweighted network, we assign edge weights using the Laplacian node embeddings,
\begin{equation}
    A[i,j] = x_i^Tx_j
\end{equation}
The edge weights then correspond to influences coming from the neighboring nodes and can be aggregated.

\subsection{Unified Propagation of Complex Contagion} 
\label{sec:prop model}
There are many variations of the complex contagion model ~\cite{centola2007complex}, among which the most commonly used is the $k$-complex contagion: a node activates when at least $k$ of its neighbors have been activated. In contrast, simple contagion models ~\cite{granovetter1978threshold,kempe2003maximizing} such as linear threshold, independent cascade and SIR can cause infections from just a single neighbor and enjoy submodularity properties. The submodularity property endows the propagation with diminishing returns, and therefore is amenable to a fast greedy approximation for downstream tasks such as influence maximization (IM). However, complex contagion is not submodular in its original form, and variants have been proposed to achieve submodularity for the ease of analysis ~\cite{gao2016general}. While these models are very expressive and have been used extensively for influence maximization, they don't account for (1) node features, and (2) cannot give rise to the characteristic S-curve in technology adoption ~\cite{christensen1992exploring} or exhibit phase transition behavior ~\cite{kleinberg2010talk}. In this work, we propose a complex contagion model that addresses the aforementioned issues and combines 3 sources of influence in a threshold based model. In the following subsections we define each influence source separately and show how they're aggregated to run propagation simulations. 

\paragraph{\textbf{Propagation Affinity}}
For a propagation $C \in [0,1]^d$ from the space of node features, we can construct a normalized vector $\hat{C}=\frac{C}{||C||}$ in the $d$-dimensional hypersphere to denote the direction of propagation. Inspired by prior work on opinion dynamics ~\cite{gaitonde2021polarization}, we define the propagation affinity for a node $v$ as the dot-product similarity between the propagation and the node feature vectors. 
\begin{equation}
    F(\hat{C},v) = \hat{C}^Tx_v
\end{equation}
Note that by construction the dot-product $\hat{C}^Tv \in [-1,1]$ since both $\hat{C}$ and $x_v$ are normalized vectors. However, to keep influences calibrated across the 3 sources, we scale the score so that it lies in $[0,1]$ as follows,
\begin{equation}
    \hat{F}(\hat{C},v) = \frac{1 + \hat{C}^Tx_v}{2}
\end{equation}

\paragraph{\textbf{Local Influence}}
The local influence corresponds to the neighborhood effects and/or the peer pressure, as also modelled by prior models of complex contagion. While most of the focus has been on the \textit{number of activating neighbors}, we argue that a better measure is the \textit{density of activating neighbors}. This has the effect, for instance, of allowing a low-degree peripheral node to activate even if the actual number of activating neighbors is very low. Explicitly, 
\begin{equation}
    LI(v,Nbr(v)) = \frac{\sum_{w \in M(v)} A_{v,w} - \sum_{w \in Nbr(v)\setminus M(v)} A_{v,w}}{d_v}
\end{equation}
where $M(v)$ denotes the set of activating neighbors of $v$ and $d_v = \sum_{w \in Nbr(v)}A_{v,w}$ is the degree of node $v$. This local influence term accounts for the competitive influence weights for activation from the neighbors and their density. Similar to the propagation affinity, we scale the value so that it lies in $[0,1]$. 
\begin{equation}
    \hat{LI}(v,Nbr(v)) = \frac{1 + LI(v,Nbr(v))}{2}
\end{equation}
The local influence term is similar in spirit to the coordination games model ~\cite{kleinberg2007cascading, easley2010networks} where nodes get payoffs from neighboring nodes for aligning their behaviors.

\paragraph{\textbf{Global Influence}}
Besides the network influences, population level influences also account for the spread of contagions ~\cite{easley2010networks}. This is especially true in OSNs, where the population level adoption statistics are provided directly to the user. Therefore we use the rate of adoption as a measure of global influence,
\begin{equation}
    GI(\hat{C},G)= \frac{|S|}{n}
\end{equation}
where $S$ is the dynamically varying set of activating neighbors. Clearly, $GI(\hat{C},G) \sim 0$ initially and increases to $1$ as $S$ grows. 

\paragraph{\textbf{Aggregating Influences and Simulating Contagion}}
Cumulatively, the 3 influence scores can be aggregated to yield the probability of activation for $v$,
\begin{multline}
    \text{Prob}(v|Nbr(v),\hat{C}, G) = \gamma(\alpha \hat{F}(\hat{C},v) + \beta LI(v,Nbr(v))\\ + (1-\alpha-\beta)GI(\hat{C},G))
    \label{eq:7}
\end{multline}
where $0 \leq \alpha \leq 1,0 \leq \beta \leq 1,$, $\alpha + \beta \leq 1$ and $\gamma$ is a parameter for modulating the simulation temperature. 

We use $\text{Prob}(v|Nbr(v),\hat{C})$ to perform Markov Chain Monte Carlo (MCMC) simulation of cascades and to draw samples from the distribution of possible propagation outcomes. In particular, we construct a Markov chain over network activation states and use a Metropolis-style update rule in which candidate activation updates are proposed according to the Bernoulli probabilities induced by $\text{Prob}(v|Nbr(v),\hat{C})$ and accepted deterministically in the irreversible activation setting.

Specifically, at each time step $t$, we propose activation updates for all currently inactive nodes independently according to their Bernoulli probabilities and accept all successful proposals. 
\[ {Y^t_{v}} = \left\{
  \begin{array}{lr}
    1 &:Y^{t-1}_v = 1, \\
    X\sim\text{Bern}(\text{Prob}(v|Nbr(v),\hat{C},G)) &:\text{otherwise} 
  \end{array}
\right.
\]
Note that once a node gets activated it stays activated for the rest of the simulation. The time steps of the simulation can be of the order of days, weeks or months in a real-world scenario, whereby nodes activate asynchronously because of multiple contacts with neighboring nodes in the network. We keep redrawing node samples until the network state hasn't changed for a minimum cooling period of $\epsilon$ steps, and this yields the steady state. Since activations are irreversible in our model, the resulting Markov chain is monotone and converges to a stationary distribution over final cascade states. This yields a valid MCMC sampler over cascade realizations induced by the propagation dynamics.




\subsection{Conditions for Local Incubation}
\label{sec:incubation}

In this section, we derive conditions for local incubation for the foregoing propagation model, primarily focusing on the second adopter.; the analysis for subsequent adopters is similar. Let the seed node $v_0$ be connected to a clique $\{v_1, v_2, \dots, v_k\}$ of size $k$. Without loss of generality, let $v_0$ attach to the clique at $v_1$. Then, from Eq.~\ref{eq:7}, the activation probability for $v_1$ is,
\begin{align}
    \text{Prob}(v_1|Nbr(v),\hat{C}) &= \gamma(\alpha \hat{F}(\hat{C},v_1) + \beta \hat{LI}(v,Nbr(v))) \\
                                    &= \gamma(\alpha \frac{1 + x^T_{v_1}\hat{C}}{2} + \beta \frac{1 + \frac{1-(k-1)}{k})}{2}\\
                                    &= \gamma(\alpha \frac{1 + x^T_{v_1}\hat{C}}{2} + \beta \frac{1}{k})
\end{align}
where we have assumed $GI(\hat{C},G))\sim0$ for a large network during local incubation and taken an unweighted network; the extension to the weighted case is straightforward. Given a cooling period of $\epsilon$ steps, the likelihood of node $v_1$ activating is the probability of success in $\epsilon$ trials of $\text{Bern}(\text{Prob}(v_1|Nbr(v),\hat{C}))$,
\begin{equation}
    \text{Prob}^\epsilon(v_1) = 1 - (1 - \gamma(\alpha \frac{1 + x^T_{v_1}\hat{C}}{2} + \beta \frac{1}{k}))^\epsilon
\end{equation}
For instance, taking $\alpha=\beta=\frac{1}{2},\gamma=1 \text{ and } \epsilon = 1$, 
\begin{equation}
    \text{Prob}^\epsilon(v_1) = \frac{k(1+x^T_{v_1}\hat{C}) + 2}{4k}
\end{equation}
Since $x^T_{v_1}\hat{C} = 1$, and assuming $k >> 2$,
\begin{equation}
    \text{Prob}^\epsilon(v_1) \sim \frac{1}{2}
\end{equation}
We can see here that even with complete alignment to the propagating notion, the probability of activation is squarely in the middle due to the singleton local influence initially. On the other hand, letting $x^T_{v_1}\hat{C} \sim  -1$, clearly $\text{Prob}^\epsilon(v_1) \sim 0$.  Increasing $\epsilon$ has the effect of increasing the the number of trials, allowing lower-probability evens to happen. For the per-step probability, we can see that when $k$ is small the second term contributes more local influence. Therefore, a smaller group favors chances of local incubation. In a larger group, the propagation affinity score needs to be higher since local influence for the bridge node will be very small. 

\subsection{Connection to Random Walks}
The proposed contagion model is connected to existing random walk diffusion models on networks. However, the proposed contagion model is more like a stochastic growth process that can be seen as a gated, time-varying random walk.  The propagation can be written in matrix-vector form precisely as,
\begin{equation}
   Y^{t+1} \sim \gamma(\alpha \hat{F}(\hat{C},V) +  \beta (2Y^t-1)D^{-1}A + (1-\alpha -\beta)GI(\hat{C}^t,G)I_n)1_n
\end{equation}
where  $\hat{F}(\hat{C},V)=\text{Diag}(\hat{F}(\hat{C},v_1),\dots,\hat{F}(\hat{C},v_n))$ is the diagonal matrix of node affinity scores, $I_n$ is the $n\times n$ identity matrix and $1_n$ is the vector of all ones. The matrix modulating the propagation is very much like a random walk matrix ~\cite{zhu2002learning,zhu2002learning,zhu2003semi}, however it's dynamic and depends on the activation states of the nodes. Moreover, the probabilistic node activations function as a gate that changes the transition matrix. Moreover, unlike a classical random walk, once activated the nodes act as constant sources of flow. Finally, there is no closed form analytical solution in terms of the system's eigenvalues for the rate of convergence due to stochasticity. Instead, we run simulations to estimate asymptotic convergence rates.

\subsection{Phase Transition Behavior}
Understanding phase transitions in networked systems is an important area of research in psychology, ecology and dynamical systems ~\cite{shrager1987observation,scheffer2012anticipating}. In the context of social contagion, this has not been studied significantly, and is especially hard because complex contagions have a distribution skewed towards small spreads, and viral events lie in the tail end of that distribution. While large-scale datasets have recently become available for this purpose ~\cite{goel2015structural}, they account more for information diffusion and not behavioural diffusion. However, it's well known from studies in management science that technological adoption follows an S-curve \cite{christensen1992exploring}, where early adopters incubate an invention and then after a slow initial period of adoption the invention reaches critical mass and propagates rapidly to a large population, with late stage adopters forming the top part of the S-curve. This can be further linked to a recent sociological study ~\cite{centola2021influencers} that shows how peripheral institutional groups can propagate large spreads through core-periphery adoption pathways. Our proposed contagion model captures this phase transition behavior and achieves periphery-core spreading. Intuitively, the contagion accelerates once the initial nodes in the PA network, which happen to have the highest centerality, get infected. Therefore, in the initial stages spreading is slow due to lower branching, and gets accelerated from both high branching through high centerality nodes and critical mass effects. Notably, our contagion model explicitly accounts for global influence and drives the propagation towards virality. This phase transition behavior is more subtle than previous works where mere changes in connectedness lead to the formation of a giant component~\cite{janson1993birth}. 




\begin{figure}[h!]
  \begin{center}
    \includegraphics[width=0.49\textwidth]{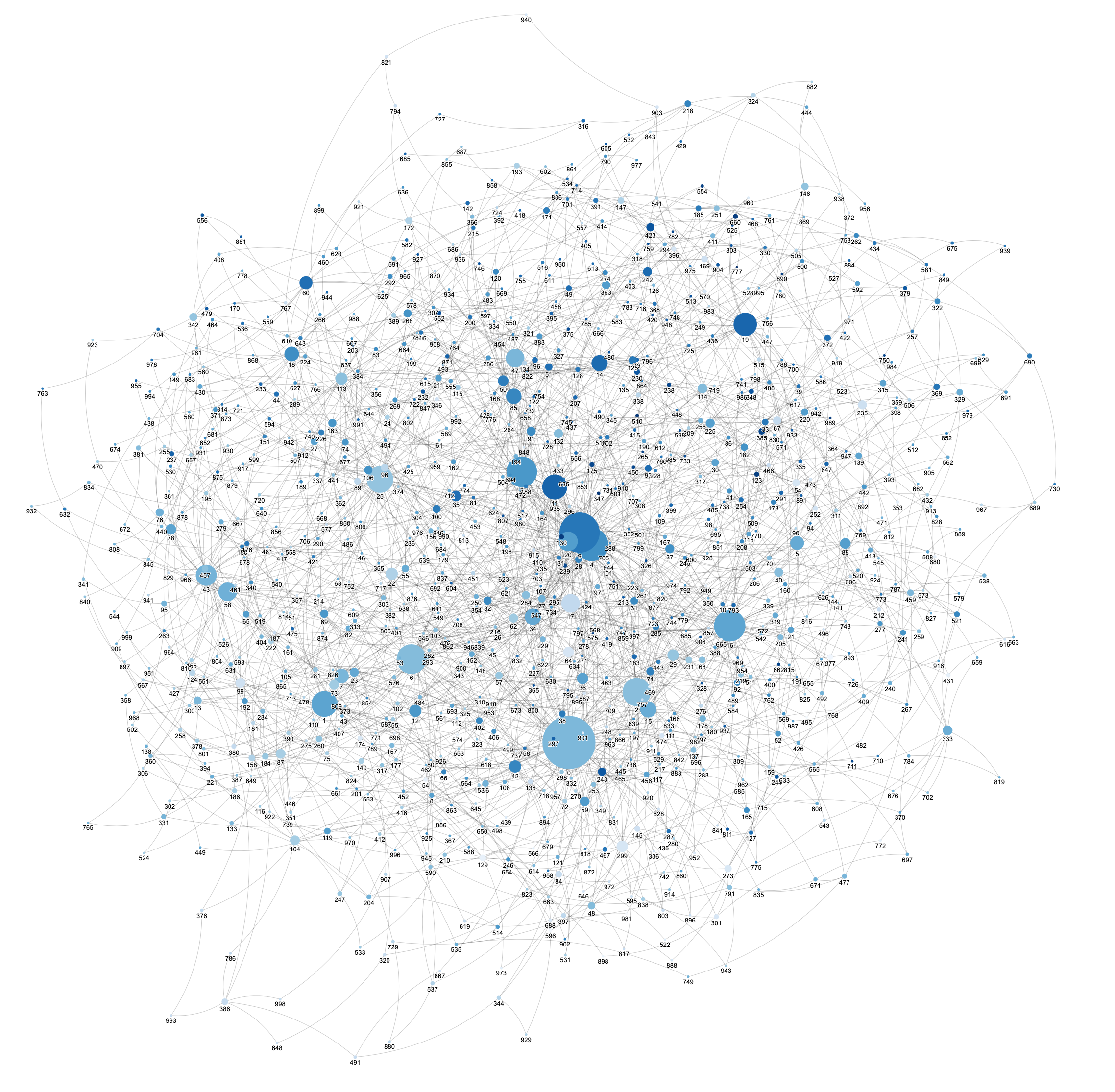}
  \end{center}
  \caption{Qualitative illustration of a PA network, $G(1000,2)$ used in our simulations. Node size and color opacity indicate degree and the component along the first spectral embedding vector respectively.}
  \label{fig:qualitative_network}
\end{figure}


\section{Experiments}

We evaluate the proposed propagation–driven cascade model on PA networks in order to analyze how stochastic propagation, reinforcement effects, and network structure jointly shape diffusion outcomes. In all experiments, we simulate influence cascades generated from the propagation dynamics described in Section~\ref{sec:prop model}, while varying the location of initial seeds, the spread control parameters, and the structural properties of the underlying networks.

Specifically, we wish to answer the following research questions:
\begin{itemize}
    \item \textbf{RQ1}: What kind of spread distribution is attained?
    \item \textbf{RQ2}: What're the dynamics of spread evolution to attain virality?
    \item \textbf{RQ3}: How does the time to attain virality vary with network structure?
    \item \textbf{RQ4}: How does the simulation vary with the parameters?
    \item \textbf{RQ5:} How do misaligned ideas get propagated?
    \item \textbf{RQ6:} How does complex contagion spread on real-world networks?
\end{itemize}

\paragraph{\textbf{Implementation details.}}
Unless stated otherwise, all experiments use the same simulation configuration. For each node, we draw $20$ stochastic samples of propagation trajectories. The propagation parameters are fixed to $\alpha = 1/3$ and $\beta = 1/3$, and the simulation temperature set to $\gamma = 0.05$. The cooling down period is set to $\epsilon=10$ number of steps. For every node, the initial propagation vector is initialized to be the node’s own feature vector; consequently, all experiments evaluate \emph{self-propagation} dynamics unless explicitly noted otherwise. The underlying network is generated using a preferential attachment (PA) model with $1000$ nodes, where each newly arriving node attaches to $2$ existing nodes. We construct node features by taking the first $k=10$ eigenvectors of the graph Laplacian. A qualitative visualization of the resulting network structure is shown in Figure~\ref{fig:qualitative_network}.  

We categorize nodes into three structural segments based on degree centrality: \emph{core} nodes correspond to the top $10\%$ of nodes ranked by degree, \emph{peripheral} nodes correspond to the bottom $10\%$, and the remaining nodes are treated as \emph{intermediate}. Throughout the experiments, we report and compare trends separately for these three node segments in order to analyze how core--periphery structure shapes propagation and cascade dynamics.

\subsection{RQ1: Spread distributions.}
\begin{figure}[h!]
    \centering
    \begin{subfigure}[t]{0.21\textwidth}
    \includegraphics[width=\textwidth]{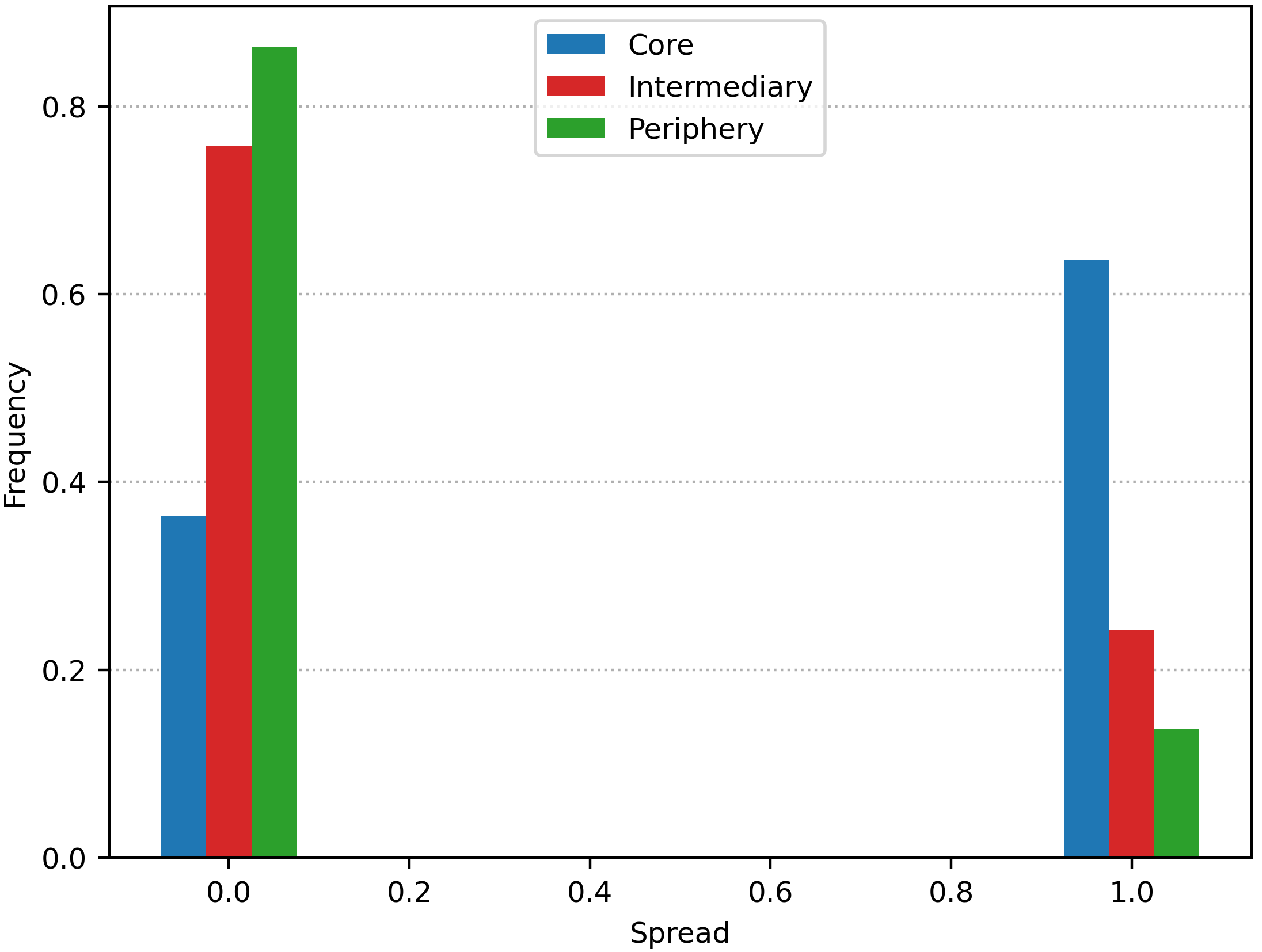}
    \caption{}
    \label{fig:spread_dist}
    \end{subfigure}
    \begin{subfigure}[t]{0.23\textwidth}
        \includegraphics[width=\textwidth]{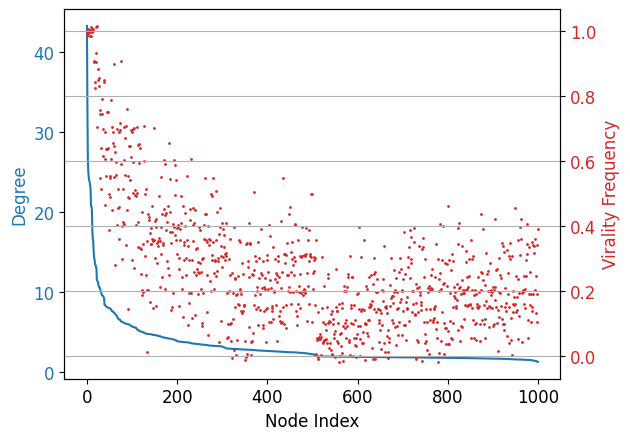} 
        \caption{}
    \label{fig:degree_to_virality_correl}
    \end{subfigure}
    \caption{(a): Distribution of final cascade sizes across simulation runs, showing a bimodal separation between localized and large-scale cascades. (b): Correlation between average seed-node degree and cascade size, illustrating the influence of core versus peripheral seeding.}
\end{figure}

In Figure~\ref{fig:spread_dist} we illustrate the distribution of final cascade sizes across repeated simulations, together with the relationship between the average degree of seed nodes and the resulting spread in Figure~\ref{fig:degree_to_virality_correl}.

The distribution in Figure~\ref{fig:spread_dist} exhibits a clearly bimodal structure. Most realizations terminate quickly after affecting only a small fraction of the network, while a smaller but significant fraction of runs lead to large-scale cascades that reach a substantial portion of the graph. This behavior is consistent with the stochastic sampling nature of our propagation model and reflects the presence of two dominant regimes: localized diffusion confined to a neighborhood of the initial seeds, and global diffusion driven by reinforcement and repeated exposure. We note not all propagations from core nodes reach virality, such as high-profile failures (eg. Google Glass) where slow initial growth leads to extinctions. Also, some propagations from peripheral nodes do reach virality, corresponding to focused groups that have core-periphery pathways to adoption~\cite{centola2021influencers}.

The degree–virality relationship shown in Figure~\ref{fig:degree_to_virality_correl} reveals a strong positive correlation between the structural centrality of the seed nodes and the probability of large-scale spread. Seeds located in the core of the PA network, characterized by high degree and dense connectivity, are substantially more likely to trigger global cascades. In contrast, cascades initiated from peripheral nodes typically remain small and short-lived. This observation aligns with the core–periphery pathways emphasized by our model: propagation originating in the core rapidly encounters multiple reinforcing paths, whereas peripheral seeds lack sufficient redundant exposure to overcome activation thresholds.

\subsection{RQ2: Spread Evolution toward Virality.}
Amongst the simulations that achieve virality, we want to understand the propagation dynamics that drive it and whether we observe a phase transition with a tipping point.

\begin{figure}[h!]
    \centering
   \begin{subfigure}[t]{0.22\textwidth}
    \includegraphics[width=\textwidth]{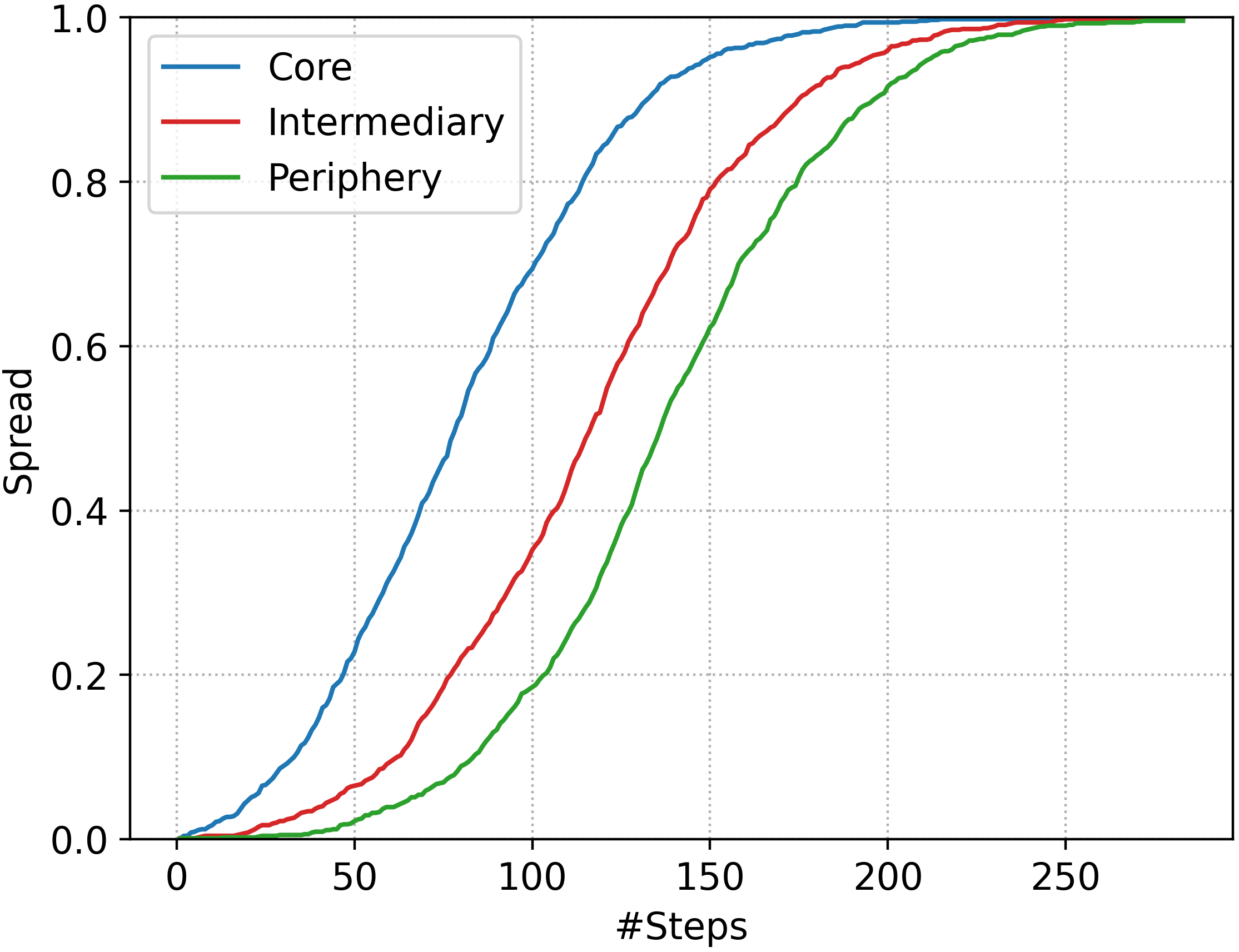}
    \caption{}
    \label{fig:spread_evolution}
    \end{subfigure}
    \begin{subfigure}[t]{0.23\textwidth}
    \includegraphics[width=\textwidth]{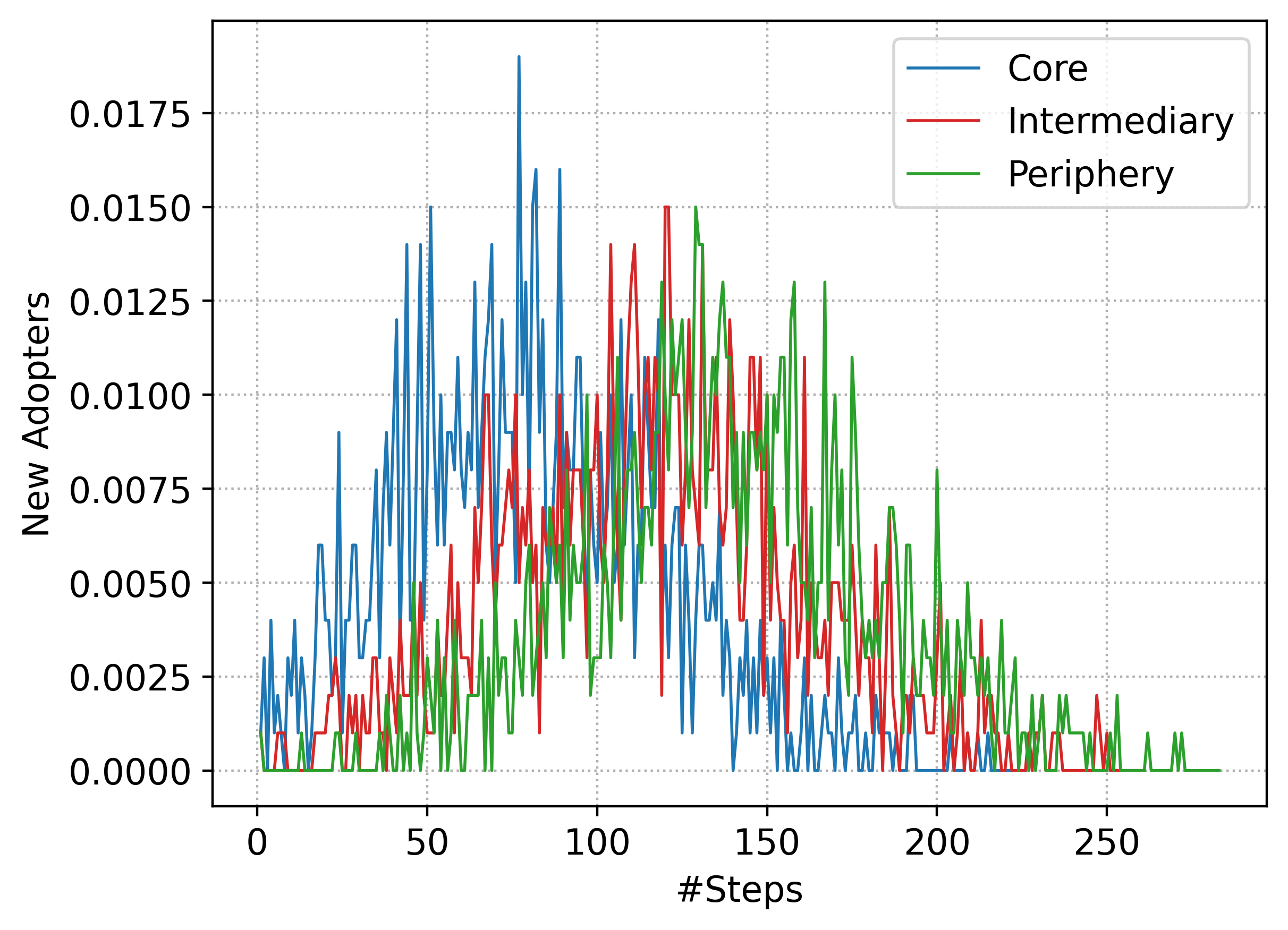} 
    \caption{}
    \label{fig:new_adopters}
    \end{subfigure}
    \caption{(a) Temporal evolution of cumulative activated nodes during cascades. (b) Number of newly activated nodes at each time step, highlighting the non-linear phase transition into virality and the tipping point into rapid diffusion.}
\end{figure}

Figure~\ref{fig:spread_evolution} shows the cumulative number of activated nodes as a function of simulation time for representative viral simulations from the three node segments. The curves reveal a two-stage diffusion pattern. In the early phase, growth is slow and sub-linear, reflecting the incubation period where branching remains low and both local and global influence remain low as well. Once the cascade reaches a sufficiently connected region of the network, typically within the core, the growth rate increases sharply. This transition corresponds to a tipping point at which there is high branching that propels the propagation towards virality.

The same phenomenon is visible in Figure~\ref{fig:new_adopters}, which plots the number of newly activated nodes at each time step. The spike in new adopters marks the onset of the supercritical regime of the cascade process. After this point, activation becomes self-sustaining due to the abundance of overlapping propagation paths and accumulated exposure. These results demonstrate that virality in our model is not driven by a single influential event but emerges from the interaction between random-walk propagation and nonlinear activation dynamics.

\subsection{RQ3: Time to Virality}
\begin{figure}[h!]
  \begin{center}
    \includegraphics[width=0.49\textwidth]{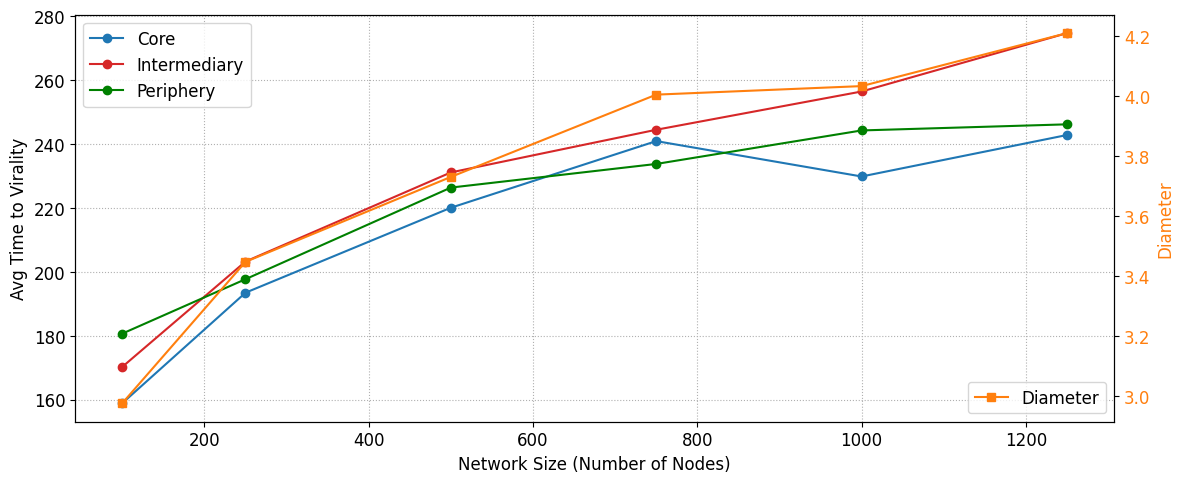}
  \end{center}
  \caption{Average time required to reach the viral regime and the network diameter as a function of network size.}
  \label{fig:time_to_virality}
\end{figure}

Figure~\ref{fig:time_to_virality} reports the average time required for a cascade to reach the viral regime as a function of network size, together with the corresponding network diameter measured on the same preferential attachment graphs. A salient observation is that the time to virality grows very slowly as the size of the network increases. Even as the number of nodes increases by a large factor, the increase in the time required to reach the viral regime remains modest, indicating a highly sublinear scaling with network size.

More importantly, the plot shows a strong correlation between the time to virality and the diameter of the underlying network. Across all network sizes, the two quantities closely track each other: instances with larger diameters consistently exhibit longer times to virality, while networks with smaller diameters reach the viral regime earlier. This tight coupling suggests that the dominant factor governing the onset of large-scale diffusion in our model is the characteristic shortest-path scale of the graph, rather than the total number of nodes. Since cascades must propagate through a sequence of intermediate activations before reaching highly connected regions that enable rapid reinforcement, the effective depth of the network---as captured by its diameter---directly constrains how quickly the process can enter the supercritical growth phase.

The observed slow growth of the time to virality is therefore consistent with known structural properties of preferential attachment graphs. In particular, the diameter of random preferential attachment networks is known to grow extremely slowly with the number of nodes, on the order of $\Theta(\log n / \log \log n)$ \cite{bollobas2004diameter}. Since propagation trajectories in our model rapidly funnel toward the network core through short paths, the number of steps required before sufficient reinforcement becomes possible remains closely tied to this slowly growing diameter. This provides a structural explanation for why the time to virality closely follows the diameter curve in Figure~\ref{fig:time_to_virality}, and only weakly depends on the overall network size.

\vspace{-1em}
 \vspace{-1em}

\subsection{RQ4: Effect of the model parameters.}



\begin{figure}[h!]
    \includegraphics[width=0.49\textwidth]{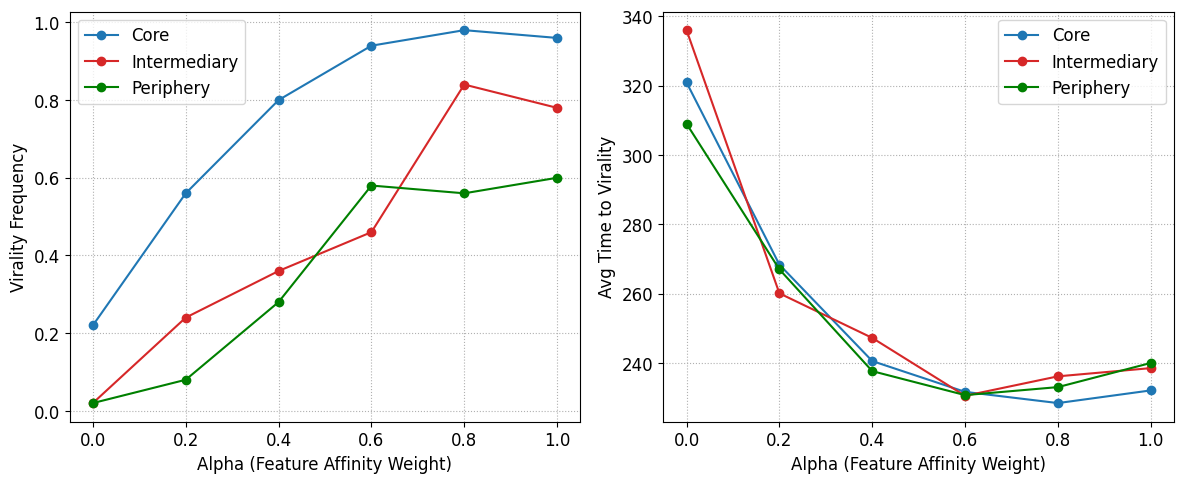}
    \caption{Effect of the feature affinity weight $\alpha$ on the virality frequency and time to virality.}
    \label{fig:alpha_sweep}
\end{figure}

\begin{figure}[h!]
    \includegraphics[width=0.49\textwidth]{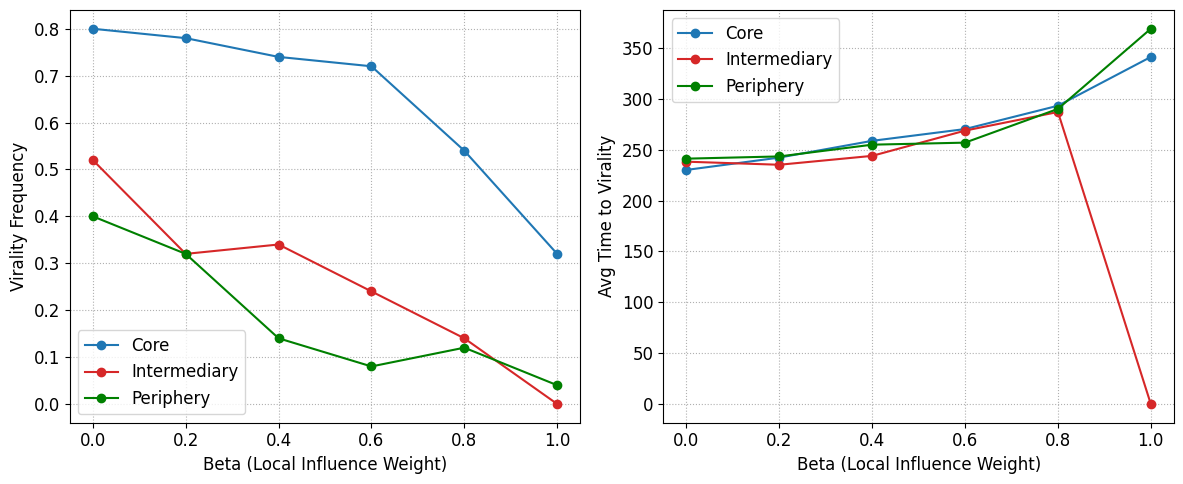}
    \caption{Effect of the local influence weight $\beta$ on the virality frequency and time to virality.}
    \label{fig:beta_sweep}
\end{figure}

\begin{figure}[h!]
    \includegraphics[width=0.49\textwidth]{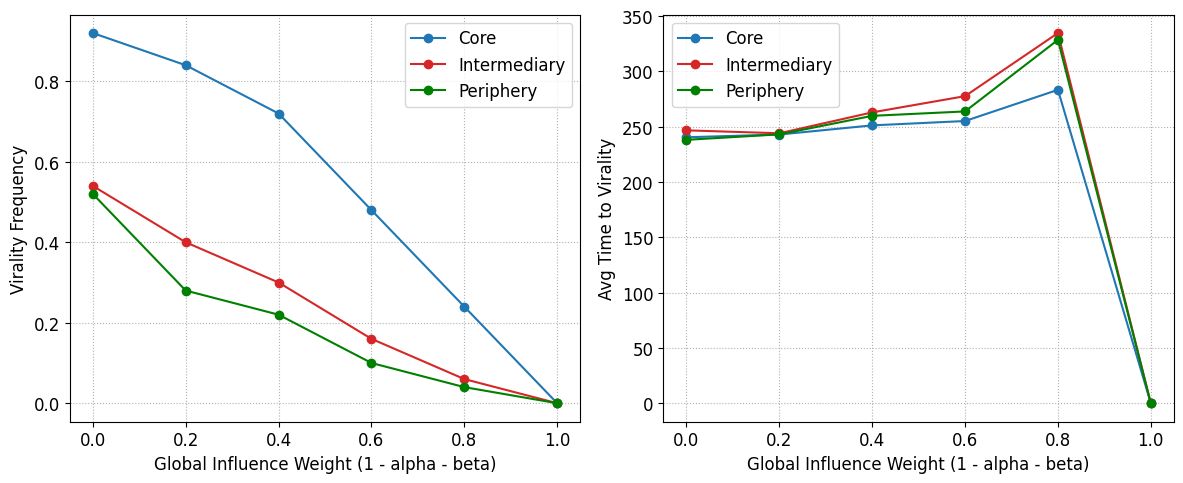}
    \caption{Effect of the global influenc weight $(1-\alpha-\beta)$ on the virality frequency and time to virality.}
    \label{fig:global_sweep}
\end{figure}

We analyze how the model parameters affect both the \emph{virality frequency}, defined as the fraction of simulation runs that enter the viral regime, and the \emph{average time to virality} among the runs that do become viral. Figures~\ref{fig:alpha_sweep}, \ref{fig:beta_sweep}, and \ref{fig:global_sweep} report these trends for the feature influence weight, the local influence parameter, and the global influence parameter, respectively.

Figure~\ref{fig:alpha_sweep} shows a monotonic increase in virality frequency together with a monotonic decrease in the average time to virality as the feature influence weight increases. This behavior is a direct consequence of the geometry of the spectral embeddings used in our experiments. The embeddings exhibit high feature similarity across large portions of the network, and as a result, self-propagating simulations are consistently reinforced by neighboring nodes with similar representations. Increasing the feature influence weight therefore strengthens early reinforcement, making it both more likely that cascades enter the viral regime and faster for successful cascades to do so.


In contrast, Figure~\ref{fig:beta_sweep} demonstrates that increasing the weight placed on local influence leads to a systematic decrease in virality frequency and smaller cascade sizes, together with an increase in the average time to virality. Strong local influence overemphasizes reinforcement within immediate neighborhoods and suppresses propagation across structurally diverse regions of the graph. This induces a network inhibition effect in which cascades become confined to locally homogeneous areas and fail to access the wide bridges required for global diffusion. This behavior is consistent with the conditions for local incubation discussed in Section~\ref{sec:incubation}: while local reinforcement is necessary to sustain early growth, excessive reliance on purely local influence prevents the transition from the incubation stage to large-scale diffusion.

Finally, Figure~\ref{fig:global_sweep} shows that increasing the global influence weight also produces a clear inhibition effect. Higher global influence reduces the virality frequency and increases the average time to virality. Strong global influence amplifies activation based on limited early exposure, effectively destabilizing the incubation stage by triggering premature activation in peripheral or weakly connected regions of the network. Because these early activations lack sufficient structural support and reinforcement pathways, many cascades fail to develop into sustained diffusion processes. Together, these results highlight that both excessive local and excessive global influence hinder the formation of successful cascades, and that a balanced interaction between local incubation and global activation is required for high virality frequency and rapid onset of virality.

\subsection{RQ5: Propagation of misaligned ideas.}

Finally, we study how ideas that are initially misaligned with the feature distributions of the population propagate in the network.

\begin{figure}[h!]
    \includegraphics[width=0.49\textwidth]{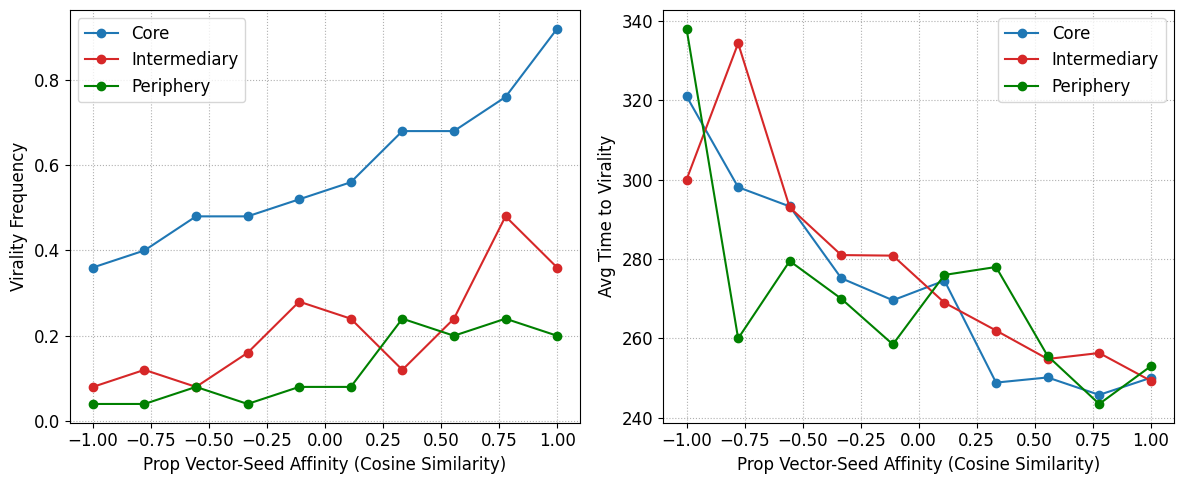}
    \caption{Effect of the propagated idea affinity to seed-node feature profiles to the virality frequency and the time to virality, illustrating the diffusion of initially misaligned ideas.}
    \label{fig:seed_affinity}
\end{figure}

Figure~\ref{fig:seed_affinity} reports the relationship between the similarity of the initial propagation vector to the feature profiles of the seed nodes and the resulting virality frequency and the average time to virality. When the propagated idea is well aligned with the local neighborhood of the seeds, diffusion proceeds smoothly and often reaches the core, enabling large and fast cascades. In contrast, ideas that are initially dissimilar to the surrounding population experience substantially lower early activation rates and times to virality, and are more likely to die before reaching structurally influential regions.

Nevertheless, a non-negligible fraction of misaligned ideas still achieve wide spread. These cases correspond to rare stochastic trajectories in which early propagation reaches highly connected nodes that provide sufficient reinforcement to overcome initial resistance. This result illustrates how structurally central actors can act as bridges for the dissemination of novel or unpopular ideas, even when local compatibility is low.

\subsection{RQ6: Real-world Complex Contagion.}

\begin{figure}[h!]
    \centering
    \begin{subfigure}[t]{0.23\textwidth}
        \includegraphics[width=\textwidth]{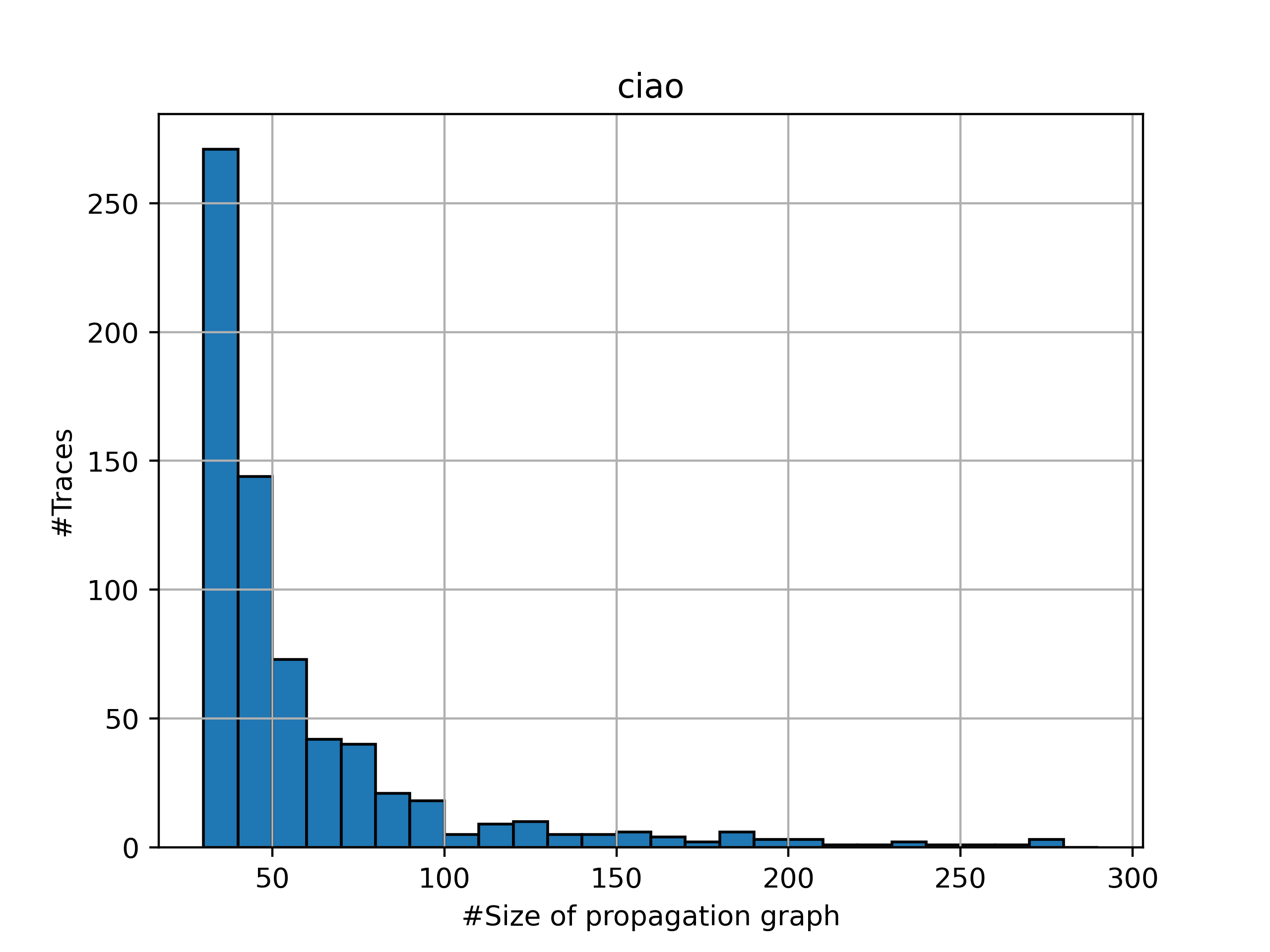}
        \caption{}
        \label{fig:}
        \end{subfigure}
      \begin{subfigure}[t]{0.23\textwidth}
        \includegraphics[width=\textwidth]{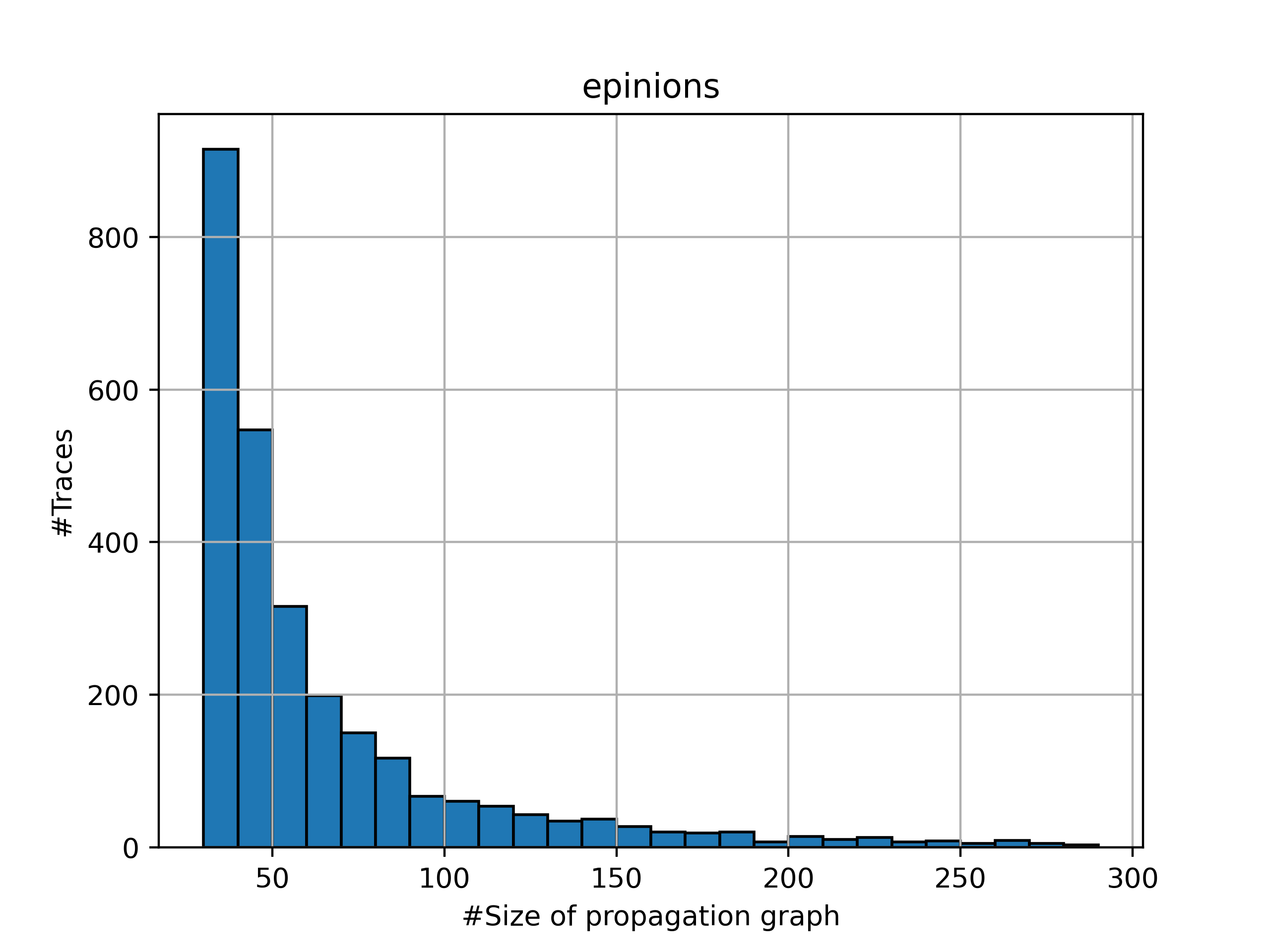} 
        \caption{}
        \label{fig:}
        \end{subfigure}
   \caption{Distribution of size of cascade traces on the (a) Ciao and (b) Epinions datasets indicates a power-law scaling where most cascades are small and there are very few big cascades.} 
    \label{fig:real_world_traces}
\end{figure}
We evaluate real-world complex contagion traces using two widely used datasets,  Epinions and Ciao, which are datasets collected from two popular product review sites, where each user can specify their trust relation in addition to rating products \cite{tang2012etrust,tang2012mtrust}. Both contain temporally ordered user interactions and trust relations, from which cascades can be reconstructed. Specifically, if  $u$ trusts $v$ and $v$ rated product $p$ before $u$ then we take it that $u$ was influenced by $v$.

A key challenge in modeling real-world complex contagion is that the empirical cascade size distributions are extremely skewed. As shown in Figure~\ref{fig:real_world_traces}, both the Ciao and Epinions datasets exhibit heavy-tailed distributions in which the vast majority of cascade traces are very small, and only a tiny fraction of cascades grow large. This strong imbalance makes learning and evaluating models of virality particularly difficult, since large cascades---which are the most informative for understanding phase transitions and global diffusion---are severely underrepresented in the data.

Moreover, real-world propagation data is substantially noisier than controlled simulations. Nodes continuously enter and leave, interaction patterns evolve over time, and multiple overlapping cascades spread simultaneously, often interfering with one another. These effects obscure the underlying propagation mechanisms and make it difficult to isolate the role of local reinforcement, global influence, and network structure. In contrast, our model provides a controlled and interpretable environment in which individual propagation trajectories can be sampled independently and the influence of specific structural and behavioral factors can be analyzed in isolation. As a result, propagation dynamics and phase transition behavior that are difficult to observe directly in real-world data become substantially clearer in the synthetic setting.

\section{Discussion}
\paragraph{Generality and feature-driven adaptability.}
A key strength of the proposed propagation--cascade framework is its generality with respect to the choice of node features and their underlying distributions. In our experiments, features are derived from spectral embeddings of the network, which induces a smooth and highly correlated feature landscape. However, the model does not rely on any specific embedding method or feature construction. In practical applications, node features may arise from specialized and heterogeneous data sources, including multimodal representations such as text embeddings, image-derived features, behavioral profiles, or contextual signals. The propagation mechanism operates directly on feature vectors and therefore naturally extends to such settings. Importantly, changing the feature distribution alters the geometry of the propagation space and the similarity structure that governs reinforcement during diffusion. As a result, different feature modalities and embedding choices can substantially modify local incubation dynamics, the accessibility of structurally diverse neighborhoods, and the likelihood of global breakout. This flexibility allows the model to be adapted to a wide range of real-world influence and information diffusion scenarios without changing the underlying propagation mechanism.

\paragraph{Early detection of phase transitions.}
Our results also suggest that the onset of large-scale diffusion can be anticipated from early-stage cascade dynamics. In particular, the growth trajectories observed during the initial incubation phase already contain strong signals of whether a cascade is likely to enter the supercritical regime. Consistent with the phase-transition behavior identified in our parameter sweeps, early growth rates and the accumulation of reinforcement paths provide a reliable proxy for the proximity of the system to its tipping point. This indicates that phase transitions in propagation behavior can be predicted at an early stage by monitoring short-term growth patterns, such as the rate of newly activated nodes and the expansion of the active frontier across structurally diverse regions. From a practical perspective, this opens the possibility of forecasting viral outbreaks or intervention opportunities before large-scale diffusion has fully unfolded.
\section{Conclusion}
We presented a propagation-driven framework for modeling complex contagion in networks that explicitly integrates feature-based affinity, local reinforcement, and global influence into a unified stochastic activation process. By representing propagating notions in the same feature space as network nodes, our model naturally captures heterogeneity in individual susceptibility and enables propagation to be interpreted as a gated, time-varying random walk.

Through extensive simulations on preferential attachment networks, we demonstrated how local incubation, core--periphery pathways, and parameter interactions jointly shape the emergence of large-scale cascades. Our results highlight that both excessive local reinforcement and excessive global influence can inhibit diffusion, and that successful cascades arise from a delicate balance between incubation and structural exploration. We further showed that phase transitions and tipping points are reflected in early-stage growth dynamics, providing opportunities for early prediction of viral outbreaks. Finally, by offering a controlled and interpretable generative framework, our model provides a complementary tool for understanding propagation mechanisms that are often obscured in noisy empirical data.


\bibliographystyle{ACM-Reference-Format}
\bibliography{main}

\appendix
\section{Comparisons to Prior Propagation Models}

Table~\ref{tab:model_comparison} summarizes the key differences between classical contagion models and our unified propagation framework. The remainder of this appendix discusses these models in greater detail and highlights how their propagation assumptions influence cascade dynamics.

\begin{table}[h!]
\centering
\small
\begin{tabular}{|l|c|c|c|c|}
\hline
\textbf{Model} & \textbf{Complex} & \textbf{Prop. Affinity} & \textbf{Tipping} \\
\hline
\textbf{IC} & No & No & Weak \\
\hline
\textbf{LT} & Yes (local) & No & Limited \\
\hline
\textbf{$k$-Complex} & Yes (hard threshold) & No & Sensitive to $k$ \\
\hline
\textbf{Data-driven} & Implicit & No & Not modeled \\
\hline
\textbf{UP (ours)} & \textbf{Yes} & \textbf{Yes} & \textbf{Strong} \\
\hline
\end{tabular}
\caption{Comparison of propagation models discussed in this paper. The unified propagation model integrates propagation affinity with local and global influence, enabling incubation dynamics and tipping-point behavior that are difficult to capture in classical contagion models.}
\label{tab:model_comparison}
\vspace{-2em}
\end{table}

\begin{figure}[h!]
    \centering
    \begin{subfigure}[t]{0.23\textwidth}
        \includegraphics[width=\textwidth]{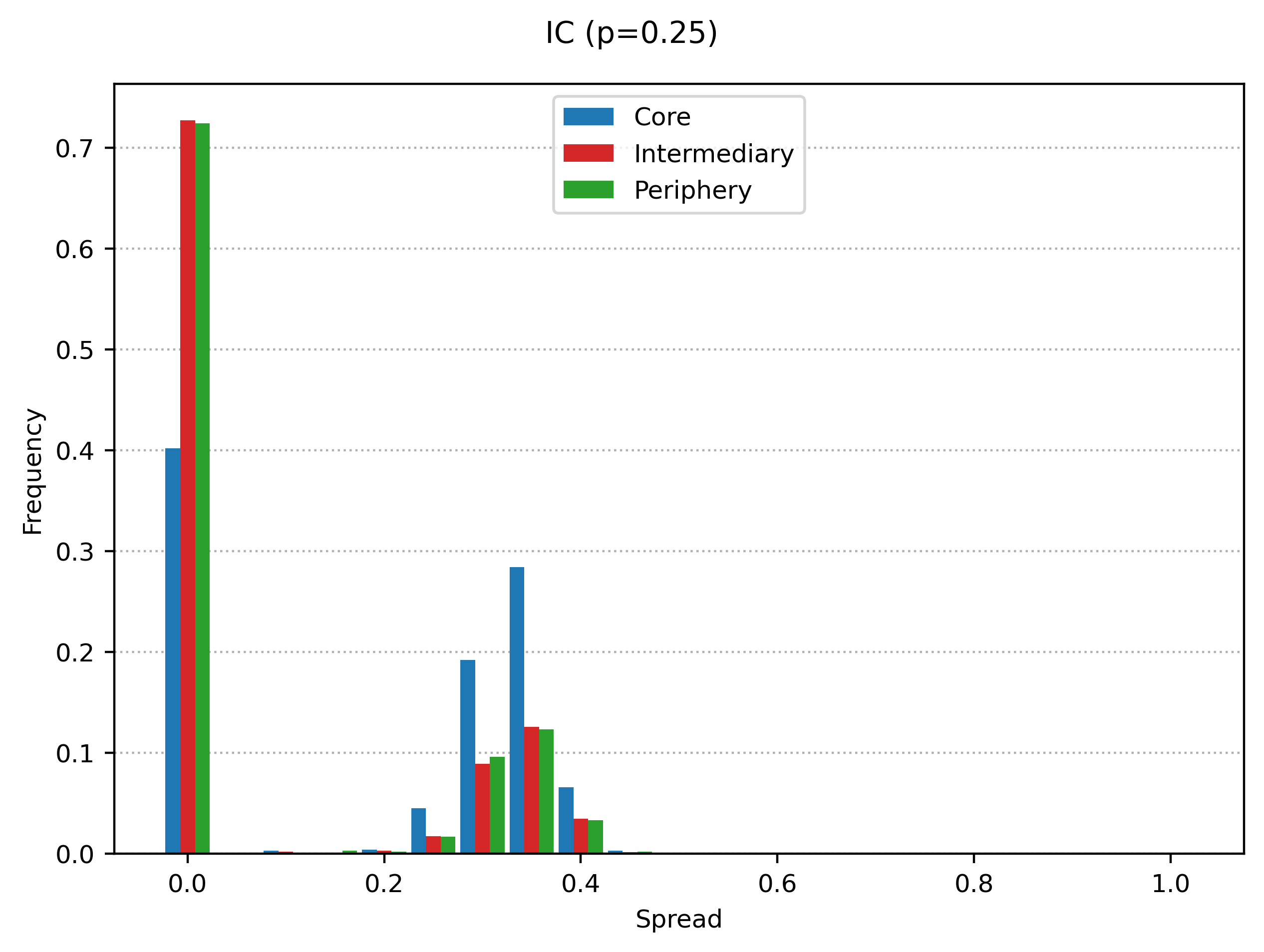}
        \caption{}
        \label{fig:}
        \end{subfigure}
      \begin{subfigure}[t]{0.23\textwidth}
        \includegraphics[width=\textwidth]{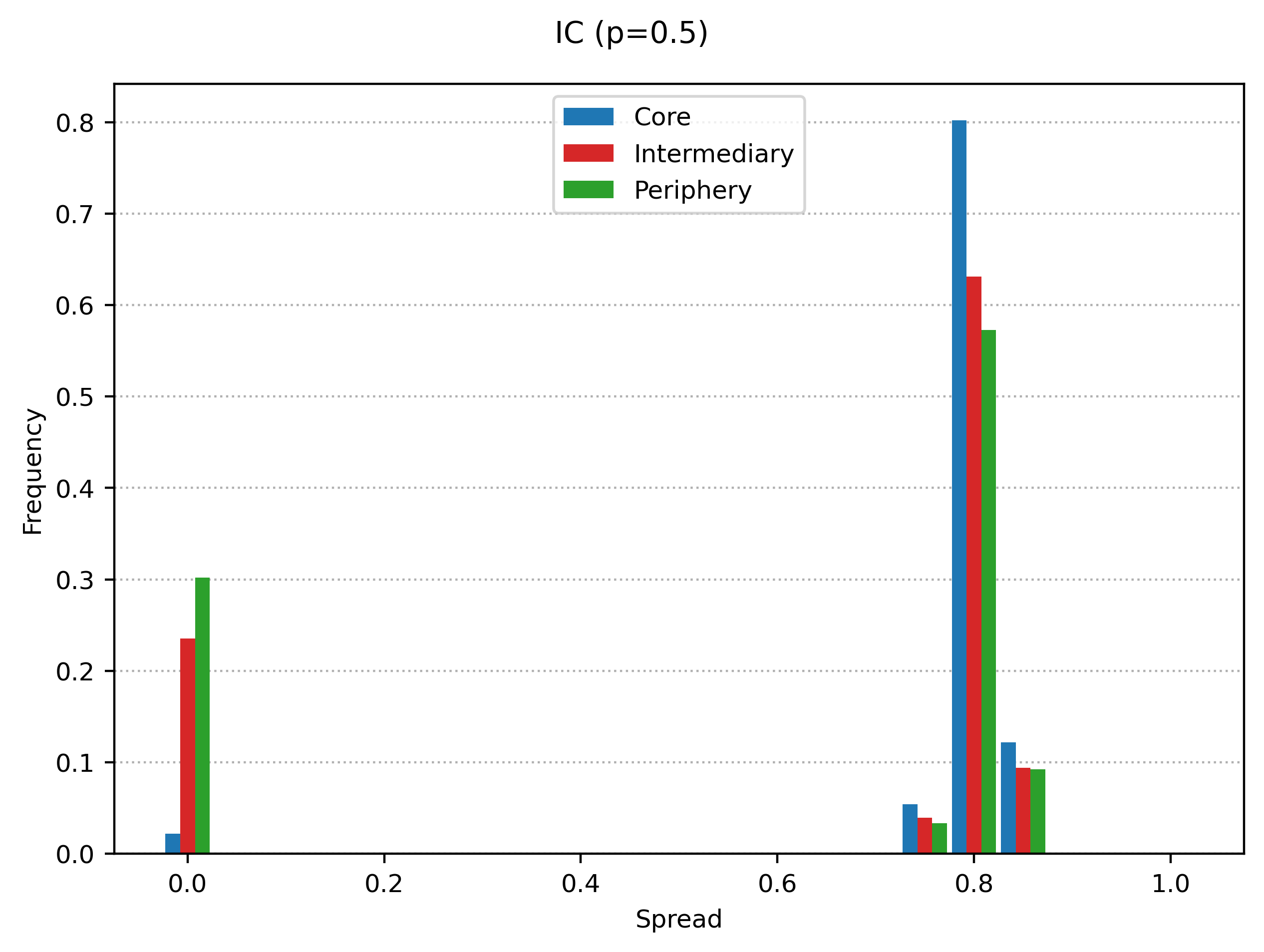} 
        \caption{}
        \label{fig:}
        \end{subfigure}
   \caption{Distribution of size of cascade traces on a 1000-node PA network for the Independent Cascade (IC) model, with activation probability (a) $p=0.25$ and (b) $p=0.5$.} 
    \label{fig:ic_spreads}
\end{figure}

\paragraph{\textbf{Independent Cascade (IC) model.}}
Figure~\ref{fig:ic_spreads} compares the cascade size distributions produced by the Independent Cascade (IC) model with activation probabilities $p=0.25$ and $p=0.5$. In both cases the IC model exhibits broad distributions with significant mass at intermediate cascade sizes. Particularly, while increasing the activation probability increases branching and shifts the distribution toward larger spreads, the resulting cascades remain concentrated around moderate spreads rather than producing the sharp tipping-point behavior observed in our unified propagation model. 

This behavior arises from the single-contact activation constraint in the IC model: each active node has exactly one chance to activate each neighbor. As a result, propagation proceeds as a branching process whose growth is limited by early stochastic failures. Increasing $p$ increases the branching factor and therefore the expected cascade size, but does not fundamentally change the qualitative shape of the distribution. Even at $p=0.5$, where branching is substantially higher, cascades do not exhibit the abrupt transition between localized diffusion and system-wide virality that characterizes our model. In contrast, the unified propagation model allows repeated reinforcement through propagation affinity and accumulated local and global influences, enabling incubation phases that can eventually trigger rapid viral growth through sufficient reinforcement.

\paragraph{\textbf{Linear Threshold (LT) model.}}
The Linear Threshold (LT) model represents another canonical contagion framework and is most closely related to the high local-influence regime of our unified propagation model. In the LT model, each node activates once the weighted influence from its neighbors exceeds a threshold drawn from a specified distribution. When these thresholds are repeatedly sampled through MCMC-style simulations, the resulting cascade realizations resemble Bernoulli sampling over activation events in our unified propagation model. However, the LT mechanism strongly emphasizes local influence and does not explicitly incorporate propagation affinity or global sentiment. Consequently, when cascades originate from a single seed, the spread distribution is typically heavily concentrated near zero. The absence of a propagation mechanism that can accumulate influence from both propagation affinity as well as local/global mechanisms prevents the formation of incubation regions capable of triggering large-scale diffusion.

\paragraph{\textbf{$k$-complex contagion.}}
The $k$-complex contagion model captures reinforcement dynamics by requiring that a node observe at least $k$ active neighbors before adopting. While this mechanism formalizes the notion of social reinforcement, it imposes strict structural constraints on diffusion. When $k=1$, the model reduces to a simple contagion and virality becomes trivial, as a single exposure is sufficient for activation. For $k>1$, however, diffusion cannot propagate from a single seed by definition, since no neighboring node can accumulate the required $k$ active contacts. As a result, the model requires either multiple simultaneous seeds or highly clustered initial conditions for diffusion to begin. In core--periphery networks this restriction makes spreading from peripheral regions particularly difficult, as peripheral nodes rarely participate in sufficiently dense local structures. In contrast, the unified propagation model allows propagation affinity in the feature space to complement structural reinforcement, enabling cascades to incubate locally and later expand through structurally advantageous pathways. This interaction between feature-driven propagation and network structure allows our model to capture tipping-point behavior that is not accessible in standard $k$-complex contagion formulations.

\section{Data-Driven Propagation Models}
While the main body of the paper focuses on analyzing propagation dynamics through controlled simulations, an important question is whether the proposed propagation framework can be learned directly from real-world cascade data. In this appendix, we explore a preliminary data-driven formulation of the unified propagation model using cascade traces from real networks. Our goal is not to build a state-of-the-art prediction model, but rather to investigate whether the propagation and activation components of the model can be estimated from empirical diffusion data and whether the resulting learned model captures meaningful propagation patterns.
\subsection{Choice of Propagation Function}   
To learn propagation dynamics from observed cascade traces, we parameterize the propagation function using a simple threshold-based architecture inspired by the Linear Threshold (LT) model.
\begin{equation}
    P(v | \text{Nbr}(v)) = \sigma(\sum_{w \in \text{Nbr}_A(v)} I_{vw} - \sum_{w \in \text{Nbr}_I(v)} I_{vw} + b_v ),
\end{equation}
\begin{equation}
    0 \leq I_{vw}, b_v \leq 0.1
\end{equation}
where $\text{Nbr}_A(v)$ and $\text{Nbr}_I(v)$ are the sets of active and inactive incoming neighbors of $v$ respectively. Here $I \in [0,0.1]^{|E|}$ and $b \in [0,0.1]^{|V|}$ are learnable parameters of the model.

Alternately, we can let the influence weights and biases be positive and otherwise be unconstrained. The initialization of the parameters is done using "small" Gaussian weights. Since the sigmoid function saturates quickly, i.e., $\sigma(5)>0.99$, it is advisable to choose mean aggregation of edge influences rather than sum aggregation as below,

\begin{equation}
    P(v | \text{Nbr}(v)) = \sigma(\frac{\sum_{w \in \text{Nbr}_A(v)} I_{vw} - \sum_{w \in \text{Nbr}_I(v)} I_{vw}}{d_v} + b_v ),
\end{equation}
\begin{equation}
    0 \leq I_{vw}, b_v
\end{equation}
where $d_v$ denotes the degree of node $v$. 

While other model architectures are possible for learning propagation dynamics from data, we adopt this formulation because it provides the closest analog to the classical Linear Threshold (LT) model. In particular, the propagation function plays a role analogous to the influence weights in LT models, while the learned activation thresholds capture heterogeneous susceptibility across nodes. This choice therefore provides a natural bridge between traditional threshold-based contagion models and data-driven learning approaches. At the same time, more expressive architectures—such as graph neural networks or attention-based propagation models—could be used to parameterize the propagation function. Our goal here, however, is to retain interpretability and maintain a clear connection with the theoretical contagion models studied in the main paper.

\subsection{Maximum Likelihood Formulation}
From Eq. 4 above, we can compute the local conditional probability $P(v|Nbr(v))$, and so compute the likelihood of a cascade $C$, which is a directed subgraph of the global graph $G$ as,
\begin{equation}
    P( C | G, I, b) = \Pi_{v \in C} P(v|\text{Nbr}(v)) \hspace{0.5em} \Pi_{v \in B_G(C)} (1 - P(v|Nbr(v)))
\end{equation}
where $B_G(C)$ is the boundary of the cascade $C$ in graph $G$, that is, the nodes in $G\setminus C$ having at least one neighbor in $C$.

We then formulate our loss function as the negative log-likelihood of all the cascades and optimize for the influence parameters $I \text{ and } b$.
Here, it is important to set the weighting ratio for the in-cascade nodes and the boundary nodes in the loss appropriately, since $|B_G(C)| >>|C|$ generally. We assign equal weight to both in-cascade nodes and the boundary nodes. 

It is also possible to differentially weigh nodes in the loss, since the early stage spreaders that're close to the seed are more important for spread prediction. Thus, we can consider subcascades within the cascade to augment our data. This can be easily implemented within the loss function.

\subsection{Prediction of Activation States}
A key question in learning propagation models from data is whether the learned model can correctly predict the intermediate activation states that arise during cascade evolution. Table~\ref{tab:activation_prediction} reports the accuracy of predicting node activation states across different node categories. We observe that the predictive accuracy for boundary nodes is consistently lower than for active non-seed nodes that lie fully inside cascade regions. 

From our deep dives into the training process we hypothesize that this behavior is due to the presence of overlapping cascades in real-world diffusion data. In datasets such as Epinions, Ciao, Twitter and APS, multiple cascades propagate simultaneously and interact with the same nodes over time. As a result, a node that lies on the boundary of one cascade may already be active due to another cascade. This makes it difficult for a single learned propagation model to accurately capture the activation behavior using deterministic thresholds and influence weights. In effect, the activation labels observed in the data may reflect the superposition of multiple propagation processes.

One possible way to address this issue would be to train models using soft labels that represent activation probabilities rather than binary states. Such labels could capture uncertainty arising from overlapping cascades and competing diffusion processes. However, learning reliable soft-label representations requires substantially larger datasets with dense temporal observations of cascade activity. In the current setting, the available data is insufficient to reliably estimate these distributions. This creates a chicken-and-egg problem: richer data would allow us to learn more expressive propagation models, but generating such data would itself require accurate generative models of cascade dynamics.

Overall, these results highlight the challenges of learning propagation dynamics directly from noisy cascade datasets and motivate the use of generative propagation models that allow controlled study of contagion mechanisms.

\begin{table}[th!]
\centering
\caption{Accuracy on prediction of activation state for active non-seeds and boundary nodes.}
\vspace{-1em}
\begin{tabular}{|c|c|c|c|}
\hline
\textbf{Dataset} & \textbf{Split} & \textbf{Active NonSeeds} & \textbf{Boundary} \\
\hline
\textbf{Epinions} & & & \\
\hline
& Train     & 0.93& 0.70 \\
\hline
& Test      & 0.62 & 0.70 \\
\hline
\textbf{Ciao} & & & \\
\hline
& Train     & 0.91 & 0.71 \\
\hline
& Test      & 0.66 & 0.68 \\
\hline
\textbf{Twitter} & & & \\
\hline
& Train     & 0.64 & 0.77 \\
\hline
& Test      & 0.54 & 0.76 \\
\hline
\textbf{APS} & & & \\
\hline
& Train     & 0.46 & 0.81 \\
\hline
& Test      & 0.34 & 0.79 \\
\hline
\end{tabular}
\label{tab:activation_prediction}
\end{table}

\subsection{Limitations of Prior Data-Driven Propagation Models}

Recent work on data-driven cascade modeling has focused primarily on predicting aggregate cascade outcomes, such as the final spread or popularity of a piece of content \cite{xu2021casflow,ling2023deep,kumar2022influence}. These models are typically trained to predict the eventual cascade size or influence spread and are often used in downstream optimization tasks such as influence maximization, where the learned models are reverse differentiated to identify optimal seed nodes.

While these approaches have achieved strong performance on spread prediction benchmarks, they do not explicitly model the intermediate activation states that define the cascade trace itself. In particular, the temporal and structural evolution of cascades—namely, which nodes activate at which stages and how propagation unfolds through the network—is not directly predicted by these models.

\begin{figure}[h!]
    \centering
    \begin{subfigure}[t]{0.23\textwidth}
        \includegraphics[width=\textwidth]{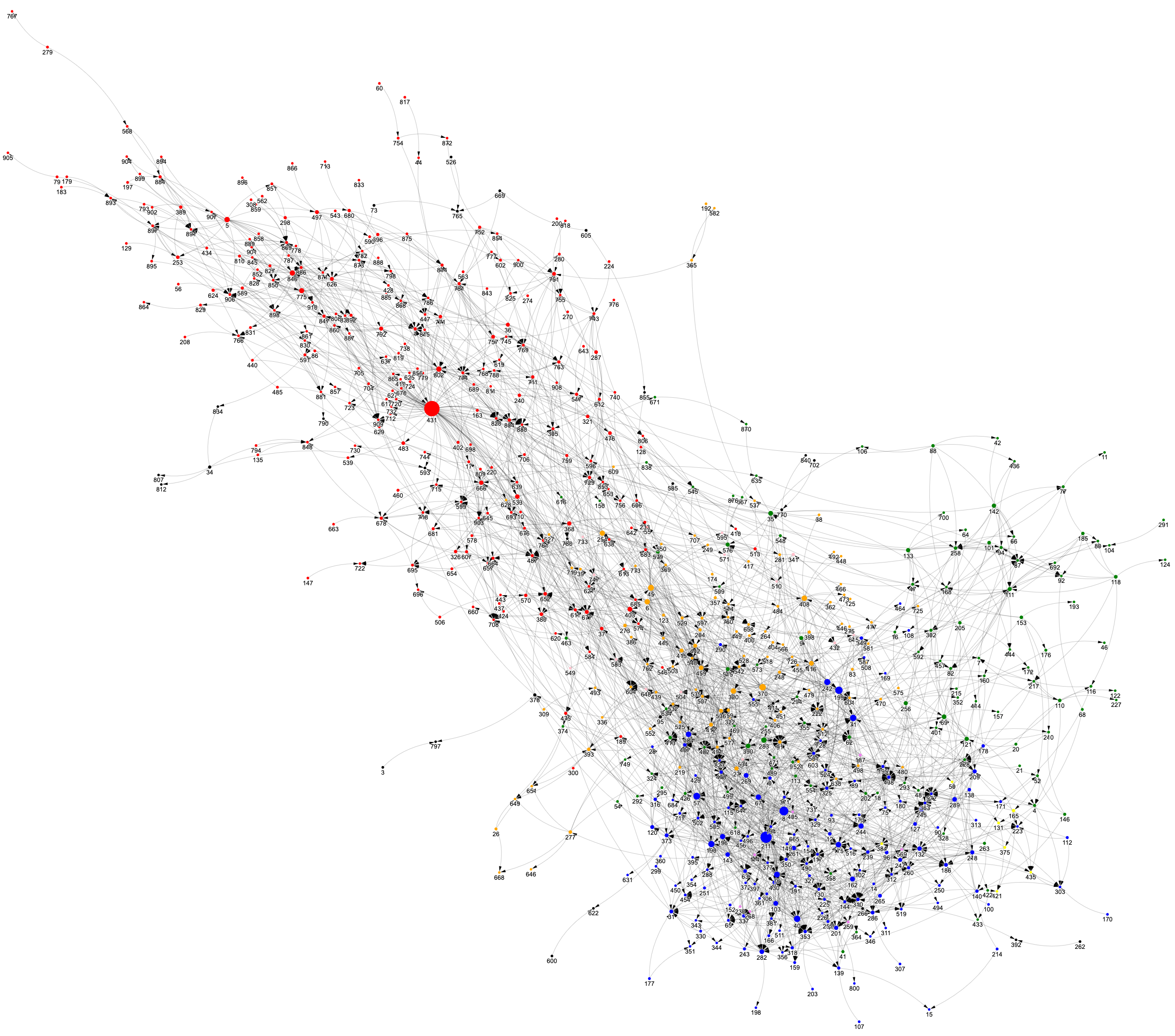}
        \caption{}
        \label{fig:}
        \end{subfigure}
      \begin{subfigure}[t]{0.23\textwidth}
        \includegraphics[width=\textwidth]{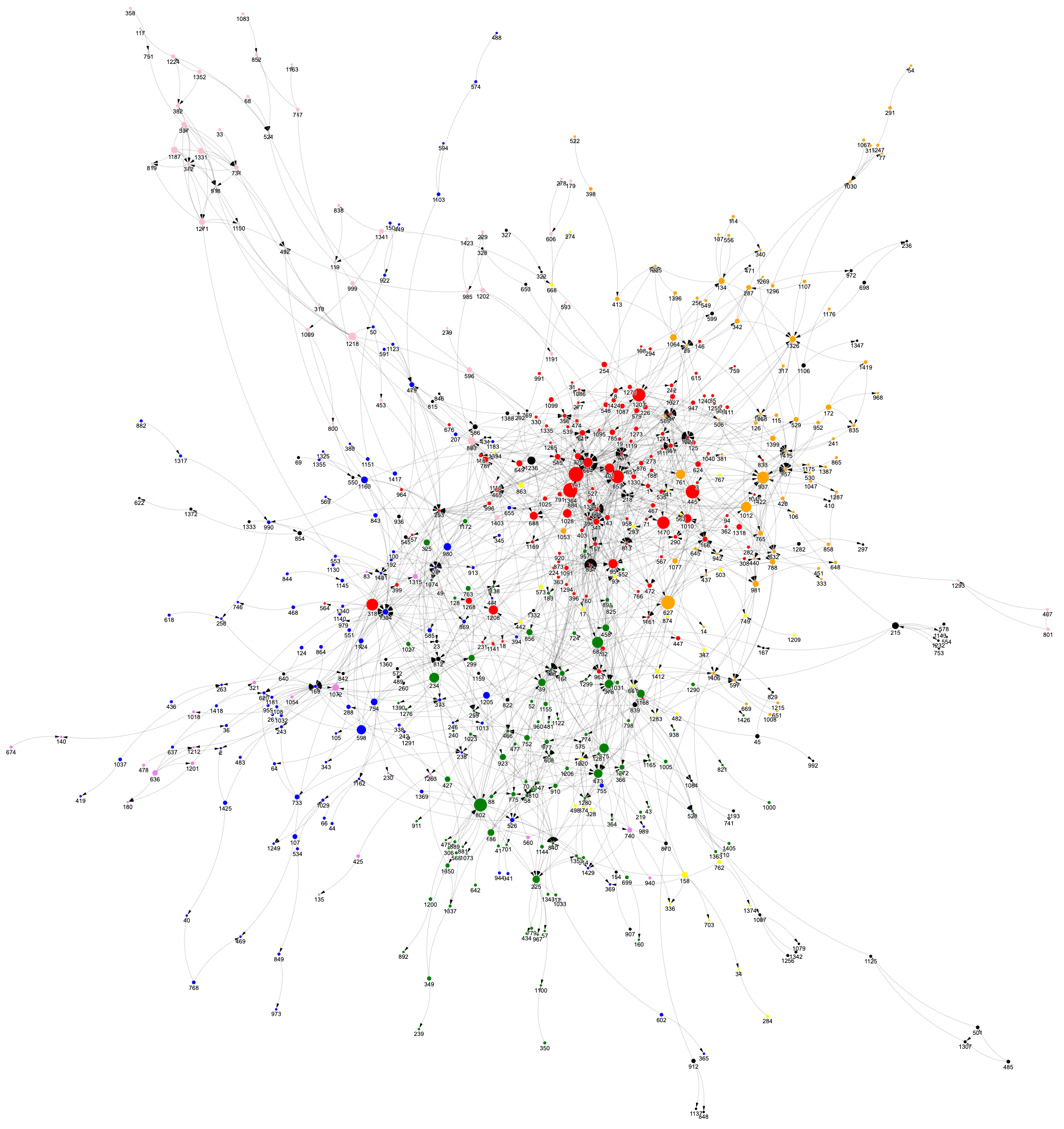} 
        \caption{}
        \label{fig:}
        \end{subfigure}
   \caption{Qualitative visualization of big cascades on the (a) Ciao and (b) Epinions datasets.} 
    \label{fig:qualitative_viz_big_cascades}
\end{figure}

However, the structure of cascade traces contains important information about the underlying diffusion dynamics. As illustrated by the qualitative visualizations of large cascade traces in the Epinions and Ciao datasets~\ref{fig:qualitative_viz_big_cascades}, the shapes of these cascades often exhibit rich structural patterns such as branching, incubation regions, and bursts of rapid expansion. Capturing these structural patterns requires models that explicitly represent the propagation process rather than only predicting final cascade sizes.

Our analysis therefore suggests that evaluation metrics for learned propagation models should incorporate structural similarity between cascade traces. In particular, graph edit distance or related structural metrics could provide a principled way to measure how closely a generated cascade matches the topology of observed diffusion traces. Such metrics would encourage models to reproduce not only the final spread but also the intermediate propagation dynamics that give rise to complex contagion behavior.

These observations further motivate the propagation-driven generative framework proposed in the main paper, which enables controlled exploration of cascade dynamics and phase transitions that are difficult to isolate in real-world diffusion datasets.

\section{Extended Propagation Model}
\label{Appendix C}

In the main formulation of the unified propagation model, node features remain fixed throughout the cascade process, while the propagation vector evolves over time through network interactions. In many real-world settings, however, activation itself changes the latent state of a node. For instance, once a user adopts an opinion, engages with a product, or participates in a social behavior, subsequent susceptibility and influence may shift accordingly. To capture this effect, we consider an extension of the model in which node features are updated after activation.

Let $\mathbf{x}_v(t) \in \mathbb{R}^d$ denote the feature vector of node $v$ at time $t$, and let $\hat{C} \in \mathbb{R}^d$ denote the propagation vector at time $t$. Suppose node $v$ activates at time $t$. We first define the unnormalized post-activation update
\begin{equation}
\tilde{\mathbf{x}}_v(t+1) = (1-\lambda)\mathbf{x}_v(t) + \lambda \hat{C},
\label{eq:feature_update_unnormalized}
\end{equation}
where $\lambda \in [0,1]$ is a tunable adaptation parameter. Since the feature vectors in our model are constrained to lie on the $d$-dimensional hypersphere, we then renormalize:
\begin{equation}
\mathbf{x}_v(t+1) = \frac{\tilde{\mathbf{x}}_v(t+1)}{\|\tilde{\mathbf{x}}_v(t+1)\|_2}.
\label{eq:feature_update_normalized}
\end{equation}
If node $v$ does not activate at time $t$, we leave its feature unchanged:
\begin{equation}
\mathbf{x}_v(t+1) = \mathbf{x}_v(t).
\label{eq:no_feature_update}
\end{equation}
The parameter $\lambda$ controls the extent to which activation shifts the latent state of a node toward the propagated notion. When $\lambda=0$, the model reduces to the static-feature setting studied in the main paper. Larger values of $\lambda$ produce stronger post-activation adaptation, moving the node more aggressively toward $\hat{C}$. Because of the normalization step in \eqref{eq:feature_update_normalized}, the updated feature remains a point on the unit hypersphere, preserving the geometric interpretation of feature similarity used throughout the model.

This update rule induces a coupled propagation--adaptation process. Since future activations depend on node features through propagation affinity, and node features themselves evolve in response to past activations, the network state becomes path-dependent over time. In particular, repeated cascades can reshape the feature geometry of the network, altering which future propagations are likely to succeed. This provides a natural mechanism for modeling \emph{concept drift} in evolving networks: the effective distribution of node states changes as a consequence of prior contagion events.

An important consequence of feature drift is that the network edge weights also evolve over time. In our setting, edge weights are derived from cosine similarity between node features. Hence, if
\begin{equation}
w_{uv}(t) = \cos\!\big(\mathbf{x}_u(t), \mathbf{x}_v(t)\big)
= \mathbf{x}_u(t)^\top \mathbf{x}_v(t),
\label{eq:dynamic_edge_weights}
\end{equation}
where the second equality uses the fact that the features are normalized, then any activation-driven update to $\mathbf{x}_u(t)$ or $\mathbf{x}_v(t)$ changes the corresponding edge weight. As a result, repeated cascades do not merely alter node susceptibilities; they also modify the effective geometry of the network itself. \textit{This yields a dynamic weighted graph in which both node states and edge weights co-evolve over time.}

This extension therefore enriches the unified propagation model in two ways. First, it allows node states to evolve endogenously as a result of activation, rather than remaining fixed throughout the process. Second, it provides a principled explanation for how repeated cascades can induce concept drift in the network over time, thereby changing both the feature geometry and the edge-weight structure under which future cascades incubate, break out, and become viral.

\section{Sampling Prop. Vectors to Maximize Spread}
\label{app:sampling_max_spread}

We now consider the inverse problem associated with the unified propagation model: given a seed node and a network, can we identify or sample a propagation vector that is likely to induce maximal spread? Let $G=(V,E)$ be a graph with adjacency matrix $A$, node feature matrix $X \in \mathbb{R}^{n\times d}$, and a single seed node $v \in V$. The goal is to produce a propagation vector $\hat{C} \in \mathbb{R}^d$ that maximizes the expected cascade spread under the stochastic dynamics of the model.

Let $\sigma(\hat{C}; v, A, X)$ denote the expected spread size of a cascade generated by initializing the unified propagation model with seed node $v$ and propagation vector $\hat{C}$. Since propagation vectors are normalized in our framework, we restrict the feasible set to the unit hypersphere. The resulting optimization problem is
\begin{equation}
\hat{C}^\star
=
\arg\max_{\|\hat{C}\|_2 = 1}
\sigma(\hat{C}; v, A, X).
\label{eq:max_spread_problem}
\end{equation}
This optimization problem is generally intractable. The search space is continuous and high-dimensional, and evaluating $\sigma(\hat{C})$ requires MCMC simulation of the stochastic cascade process. Furthermore, the objective function is highly nonconvex because small changes in $\hat{C}$ can induce activation threshold crossings that significantly alter cascade trajectories. We therefore consider approximate sampling-based strategies for identifying high-quality propagation vectors.

\subsection{Candidate generation}

A natural heuristic is to restrict the search to propagation vectors aligned with node features that lie along structurally favorable diffusion pathways. In preferential attachment networks and other core--periphery structures, viral cascades often occur when propagation reaches highly connected core nodes or traverses wide bridges between communities.

Let $\mathcal{N}_K(v)$ denote a candidate node pool constructed from

\begin{enumerate}
\item the seed node $v$,
\item nodes within $K$ hops of $v$,
\item high-degree nodes reachable from $v$, and
\item nodes lying on short paths from $v$ to the network core.
\end{enumerate}

From this pool, we construct candidate propagation vectors by normalizing node feature vectors or local feature averages:
\begin{equation}
\hat{C}_u = \frac{\mathbf{x}_u}{\|\mathbf{x}_u\|_2},
\qquad
\hat{C}_u =
\frac{\mathbf{x}_u + \sum_{w\in N(u)} \mathbf{x}_w}
{\left\|\mathbf{x}_u + \sum_{w\in N(u)} \mathbf{x}_w\right\|_2},
\end{equation}
for $u \in \mathcal{N}_K(v)$.

This reduces the search space from the continuous hypersphere to a finite set of feature-informed directions that likely align with the feature geometry of nodes capable of sustaining propagation.

\subsection{Monte Carlo estimation of spread}
For a candidate propagation vector $\hat{C}$, we estimate the expected spread by repeated cascade simulations. Let $\sigma^{(m)}(\hat{C})$ denote the spread size observed in the $m$-th cascade simulation. The empirical estimate of expected spread is
\begin{equation}
\widehat{\sigma}(\hat{C})
=
\frac{1}{M}
\sum_{m=1}^{M}
\sigma^{(m)}(\hat{C}).
\label{eq:empirical_spread}
\end{equation}
Candidate propagation vectors can be ranked according to $\widehat{\sigma}(\hat{C})$.

\subsection{Beam search refinement}
To improve over simple enumeration, we perform local refinement around high-scoring propagation vectors. Starting from a candidate $\hat{C}$, we generate perturbed candidates of the form
\begin{equation}
\hat{C}' =
\frac{\hat{C} + \epsilon \mathbf{z}}
{\|\hat{C} + \epsilon \mathbf{z}\|_2},
\end{equation}
where $\mathbf{z}$ is sampled from an isotropic noise distribution and $\epsilon$ is a small perturbation parameter. Each perturbed vector is evaluated using the Monte Carlo estimator in \eqref{eq:empirical_spread}, and the top $B$ candidates are retained for further refinement. Repeating this process for $T$ rounds yields a beam-search approximation over the hypersphere of propagation vectors.

\subsection{Dynamic programming approximation}
Although exact dynamic programming over propagation vectors is infeasible, a coarse approximation can be constructed by discretizing the propagation space into a finite codebook $\{\hat{C}_1,\dots,\hat{C}_R\}$ and grouping nodes into structural states such as \emph{core}, \emph{intermediate}, and \emph{periphery}.

Let $V_t(r,s)$ denote the estimated future spread obtainable at time $t$ when the propagation vector corresponds to codebook entry $r$ and the current cascade frontier lies primarily in structural region $s$. An approximate Bellman recursion can then be written as
\begin{equation}
V_t(r,s)
=
\max_{r' \in [R]}
\left\{
\mathrm{Reward}(r,s,r')
+
\sum_{s'} P(s' \mid s,r,r') V_{t+1}(r',s')
\right\},
\end{equation}
where $\mathrm{Reward}(r,s,r')$ estimates the immediate increase in spread and $P(s'|s,r,r')$ represents transition probabilities between structural regions estimated from simulation.

\subsection{Approximate algorithm}
In practice, the following approximate procedure can be used:

\begin{enumerate}
\item Construct candidate propagation vectors from the features of nodes near the seed $v$ and along short paths to core nodes.
\item Estimate the expected spread $\widehat{\sigma}(\hat{C})$ for each candidate using Monte Carlo cascade simulations.
\item Refine the highest-scoring candidates using beam search with normalized perturbations.
\item  Use the coarse dynamic programming approximation to prioritize candidates likely to move the cascade toward structurally advantageous regions of the graph.
\end{enumerate}

This procedure yields an approximate solver
\begin{equation}
\textsc{SamplePropagationVector}(v,A,X) \rightarrow \hat{C},
\end{equation}
which outputs a normalized propagation vector predicted to achieve large expected spread. By combining structural heuristics with Monte Carlo evaluation, the algorithm exploits the key mechanisms of the unified propagation model: propagation affinity in feature space and structural pathways that enable incubation and virality.

\end{document}